\documentclass[12pt]{iopart}

\usepackage{bm}
\usepackage{color}
\usepackage{graphicx}

\usepackage{iopams}  
\begin{document}

\title[Hypothesis on the nature and origin of cold dark matter]{Hypothesis on the nature and origin of cold dark matter}

\author{Roman Schnabel}

\address{Institut f\"ur Laserphysik \& Zentrum f\"ur Optische Quantentechnologien, Universit\"at Hamburg, Luruper Chaussee 149, 22761 Hamburg, Germany}
\ead{roman.schnabel@uni-hamburg.de}
\vspace{10pt}
\begin{indented}
\item[] June 14, 2020; October 03, 2021 with editorial changes
\end{indented}

\begin{abstract}
One of the greatest mysteries in astrophysics and cosmology is the nature and the origin of cold dark matter, which makes up more than 84\% of the mass in the universe. Dark matter reacts to gravitational forces, creates them, and determines the dynamics of stars around galactic centres, but does not absorb or emit electromagnetic radiation. 
To date, no association with known types of matter has been conclusive and no proposed new dark matter particles have been found.
Here I propose and discuss how dark matter evolved from ultra-light fermionic particles that decoupled from the rest of the universe shortly after the Big Bang. 
My description explicitly takes into account their interference and their conversion into massive dark-matter quantum fields of cosmic size.
I also argue that dark matter and supermassive black holes have the same origin and evolved at the same time, but are nevertheless statistically uncorrelated.
If the particles' decoupling time was about half a second after the Big Bang, my hypothesis predicts a minimum mass for super\-massive black holes that fits well to the smallest known such object of $5 \cdot 10^4$ solar masses. It seems very much likely that the ultra-light fermionic particle was the neutrino.
\end{abstract}

\section{The dark-matter mystery}
\begin{figure}
	\centering
	\vspace{1mm}
	\includegraphics[width=0.66\textwidth]{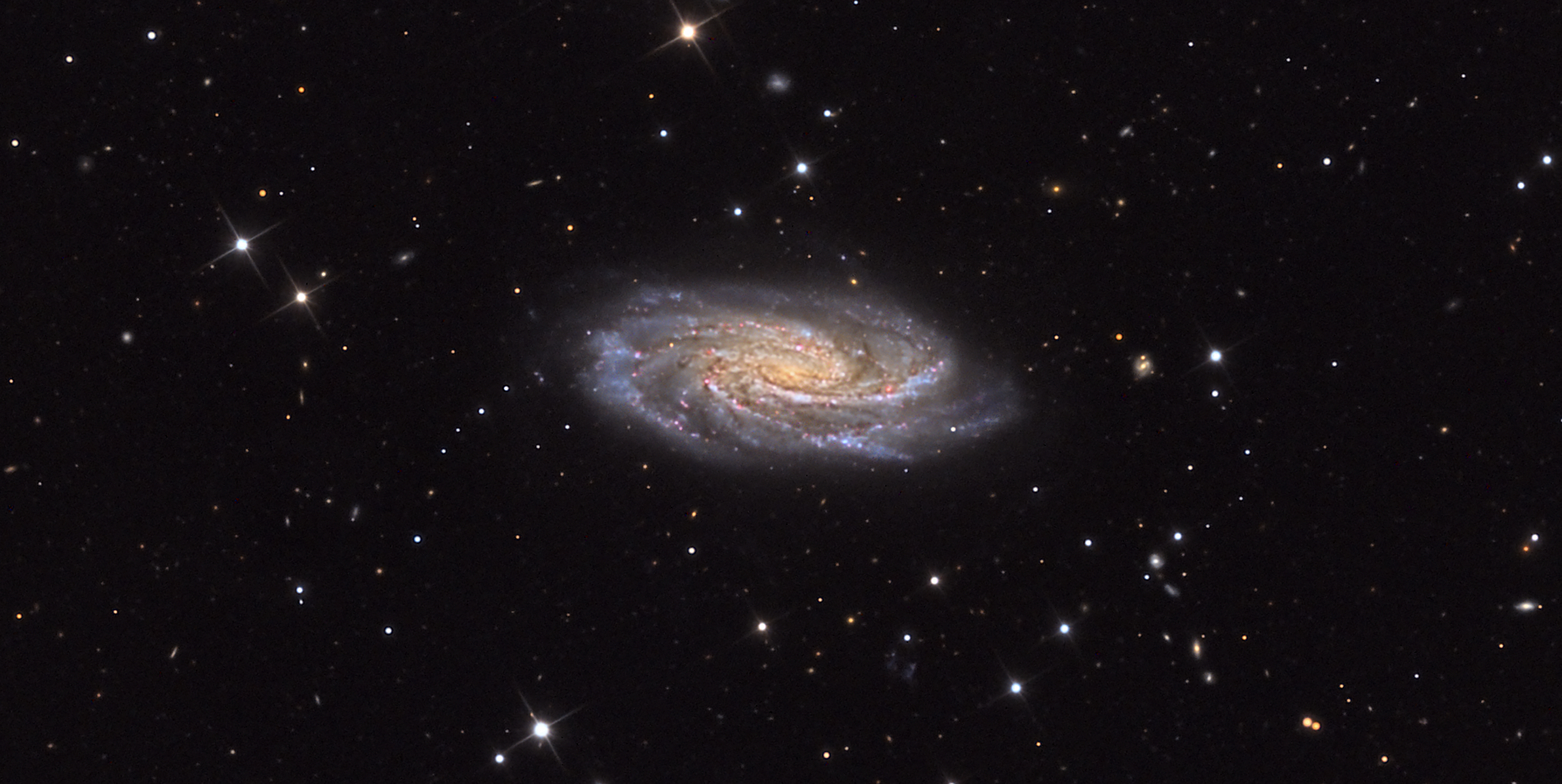}
	\vspace{0mm}
\caption{Photograph of the spiral galaxy NGC3672. Stars in the spiral arms at distances from the galactic center between a third and the full galactic radius have roughly the same speeds, as V. Rubin observed in the 1970s \cite{Rubin1978}. The fact that the speeds do not decrease at greater distances from the galactic centre can be explained by the existence of a largely extensive halo of an invisible mass of unknown nature, called \emph{cold dark matter}. Credit: Adam Block/Mount Lemmon SkyCenter/University of Arizona.}
\label{fig1}
\end{figure}

The existence of dark matter has become evident from a large number of astronomical observations that call for much more gravitating matter than what can be seen with any type of telescope. As early as the 1930s, it was concluded from the movements of galaxies in galaxy clusters that most of the matter in the universe could be dark \cite{Massey2010,DeSwart2017}. 
More recent observations of the rotational properties of galaxies confirm this \cite{Rubin1970,Ostriker1973,Einasto1974,Rubin1978,Rubin1980}, cf.~Fig.~\ref{fig1}, and suggest that the mass content of dark matter in galaxies is about ten times higher than that of the stellar components in the galactic disk and bulk \cite{Kafle2014}. The existence of dark matter was also supported by the strength of gravitational lensing by galaxy clusters. \cite{Massey2010,Mellier1999,Clowe2006}. 
Its existence is also necessary to explain the current structure of the spatial distribution of stars, galaxies and galaxy clusters, as well as the anisotropy of the cosmic microwave background (CMB).

The CMB was formed 380,000 years after the Big Bang, when protons, other light nuclei, and electrons combined to form neutral atoms. Only then could light spread freely because it was no longer continuously scattered by the charged matter. In return, the repelling radiation pressure forces between the baryonic particles were greatly reduced and the atoms began to clump together forming seeds for stars. 
However, computer models have shown that the time since the release of the CMB was too short for baryonic seeds to grow into the cosmic structure of the galaxies and galaxy clusters observed. 
In fact, the anisotropy of the cosmic microwave background, as observed by the satellite missions \emph{COBE} \cite{Smoot1992}, \emph{WMAP} \cite{Hinshaw2009}, and \emph{Planck} \cite{Ade2016} proves that matter had already formed seeds much earlier than 380.000 years after the Big Bang. This calls for (cold) dark matter that is completely invisible, i.e.~does not emit, absorb or reflect any kind of electromagnetic radiation.

\begin{figure}
	\centering
	\vspace{0mm}
	\includegraphics[width=0.4\textwidth]{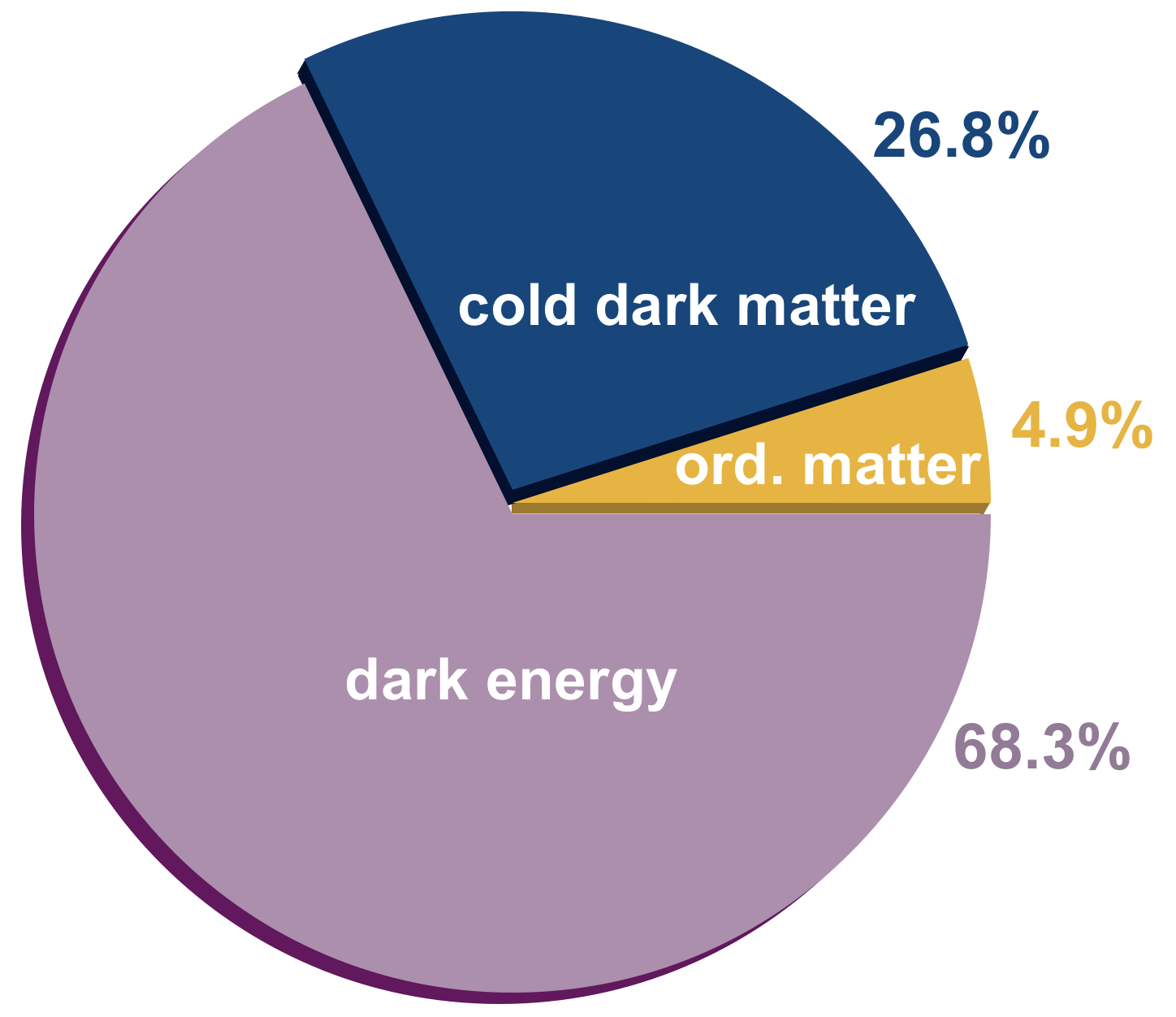}
	\vspace{0mm}
	\caption{Energy densities as they contribute to the total energy density of the universe according to the $\Lambda$CDM model and the experimental data of the Planck mission \cite{Ade2016}. Cold dark matter (`CDM') contributes 26.8\%, ordinary (baryonic) matter just 4.9\%. Cold DM thus represents 84.5\% of the gravitating energy density. The remaining 68.3\% are attributed to an unknown energy that causes the observed accelerated expansion of today's universe and is called `dark energy', described by the symbol `$\Lambda$'. Photons, ultra-relativistic neutrinos, black holes as well as gravitational waves do not contribute significantly.}
\label{fig2}
\end{figure}

A highly successful model that reproduces the evolution of the universe in consistency with astronomical observations is the `Lambda Cold Dark Matter' model ($\Lambda$CDM model) \cite{Spergel2015}. In addition to a number of fixed parameters, the model uses six \emph{free} parameters, whose values are deliberately set to achieve consistency. Amongst these are the \emph{age of the universe} with a value of about 13.8 billion years, the amount of \emph{cold dark matter} (DM), and the amount of ordinary, \emph{baryonic} matter \cite{Ade2016}, see Fig.~\ref{fig2}.
Although successful, the model does not give any further hints concerning the nature of dark matter.\\
Candidates for the particle of cold dark matter should have non-zero rest mass, be neutral but not baryonic, and be stable over cosmic time scales. Additionally, there is an argument why candidates shouldn't be light fermions: 
According to the Pauli exclusion principle \cite{Pauli1925}, identical fermions cannot occupy the same space. 
The volume that a single fermion occupies is equal to its full-width position uncertainty cubed.
The fermionic character thus limits the number density of fermions. So there is a minimum mass per fermion in order to achieve the observed mass density of dark matter. \cite{Tremaine1979}. 
There is another argument why candidates, at least if they were in thermal equilibrium with the primordial plasma sometime shortly after the Big Bang, should not have too low rest masses.
Their average speeds would be very high even at moderate temperatures, so that the gravitational structure would be washed out immediately.
These arguments basically rule out all known particles as DM candidates, including neutrinos \cite{Tremaine1979,White1983}.
The search for novel particles began many years ago, but has so far not been successful  \cite{Goodman1985,Jungman1996,Ringwald2012,Klasen2015,Bertone2018}.

\section{Proposition for solving the dark-matter puzzle}
In this article, I present the hypothesis that ultra-light fermions that were in thermal equilibrium with the hot primeval plasma \emph{cannot} be ruled out as cold DM candidates. 
I claim that such fermions are actually strong candidates for cold dark matter.
First, I outline the mechanism by which ultra-light fermions that have interacted with the hot plasma in the past naturally turn into macroscopic cold quantum fields once they stop interacting (or --~in other words~-- `decouple').
Second, I argue that two such fermions combine to form pairs of zero spin and zero momentum, similarly to electron pairs in superconductors. 
Third, I show that the excitation of these quantum fields naturally reaches occupation numbers at which they have to transform into (supermassive) black holes, around which the remaining quantum fields can aggregate as cold DM halos.
Specifically, I propose and discuss the hypothesis that cold DM actually emerged from the almost massless and fermionic neutrinos when they decoupled half a second after the Big Bang.
In the following sections I discuss the neutrino as the DM particle. 
My conclusions, to the best of my knowledge, are consistent with all astronomical observations and all major constraints that have been worked out so far.

\subsection{Neutrinos before and during their decoupling} \label{ssec:2.1}
The purpose of this subsection is to summarize known facts about the neutrinos (i) when they were in thermal equilibrium with the primordial plasma and (ii) when they began to decouple from it.
The event of neutrino decoupling is a well-established fact. It took place about half a second after the Big Bang ($t_{\nu {\rm d}} = 0.5$\,s), when the temperature of the universe was about $T_{\nu {\rm d}} \approx 3 \cdot 10^{10}$\,K \cite{deSalas2016}. 
The term `decoupling' describes a rather instantaneous process, after that the interaction rate between a neutrino and any other particle was basically zero. Neutrino decoupling was caused by the expansion of the universe and the associated reduction of energy density and temperature \cite{Sciama1994}.
At decoupling temperature $T_{\nu {\rm d}}$, the (weak-force) interaction rate of the neutrinos ($\propto T^{\;\!5\!}$) dropped below the relative one-dimensional expansion rate of the universe ($\propto T^{\;\!2\!}$), which meant that the interval between two momentum-changing interactions started to exceed the age of the universe \cite{Sciama1994}. 

Before decoupling, neutrinos were mainly coupled to the primeval plasma via the annihilation of charged leptons. For instance, neutrinos with electron flavor ($\nu_{\rm e}$) were coupled via the annihilation and creation of electron\,(${\rm e}^-$)\,/\,positron\,(${\rm e}^+$) pairs \cite{Sciama1994} 
according to  
\begin{equation}
\label{eq:chain}
\hspace{2cm}         \nu_{\rm e} + \bar{\nu}_{\rm e}\;\; \leftrightarrow \;\; {\rm e}^- + {\rm e}^+ \;\; \leftrightarrow \;\; \gamma + \gamma \; , 
\end{equation}
where $\bar{\nu}_{\rm e}$ is the electron anti-neutrino. (Note, neutrino and anti-neutrino might be identical, however, this is currently unknown \cite{EXO-200-2014}). 
Electrons and positrons, in turn, were in equilibrium with the annihilation and creation of photons ($\gamma$). 
Before neutrino decoupling, energy was equally distributed over the degrees of freedom in the chain above. Neutrino pairs, electron/positron pairs, and photon pairs were just different coexisting forms of energy. 

Before and at decoupling, the neutrinos' average thermal kinetic energy was much higher than the energy related to their tiny rest mass. Neutrinos were ultra-relativistic. In this case, the expression for the relativistic kinetic energy together with the equipartition theorem yield an average absolute neutrino momentum in three dimensions of $\langle |\vec{p}\,| \rangle_{\rm th} \approx 3 k_B T / c$, where $k_B$ is the Boltzmann constant and $c$ the speed of light. 
During the decoupling, i.e. without interactions, the neutrinos were basically a relativistic ideal fermionic gas.
As a well-known fact, $\langle |\vec{p}\,| \rangle_{\rm th}$ is then also a good approximation for the full width (fw) of the Fermi-Dirac momentum \emph{distribution}, i.e.~$\langle |\vec{p}\,| \rangle_{\rm th} \approx \Delta^{\rm fw}_{\rm th} |\vec{p}\,| $, see for instance Fig.\,1 in Ref.\,\cite{Liu2012}.\\ 

Before and during the decoupling, the positional uncertainty of the neutrino is quantified by the thermal De\,Broglie wavelength. 
For one dimension it reads
$\lambda_{\rm th} = h/ \langle |p_x| \rangle_{\rm th}$, where $h$ is the Planck constant and $\langle |p_x| \rangle_{\rm th}$ the one-dimensional average relativistic momentum. Using the relations above, one gets for the thermal De\,Broglie wavelength of ultra-relativistic particles in one dimension
\begin{equation}
\label{eq:lambda}
\hspace{2cm}        \lambda_{\rm th} \approx hc/(k_B T) \approx h/ \Delta^{\rm fw}_{\rm th} p_x \; .
\end{equation}
%
Interestingly, the product $\,\lambda_{\rm th} \cdot \Delta^{\rm fw}_{\rm th} p_x \approx h$ does not dependent on temperature. 
This fact suggests that the product represents states with minimal uncertainty (`pure' states) in the phase space spanned by position and momentum, because every additional contribution to a product with minimal uncertainty should reveal itself by a monotonically increasing function of the temperature.
The same fact suggests that not only $\,\lambda_{\rm th}$ but also $\Delta^{\rm fw}_{\rm th} p_x$ is entirely given by quantum uncertainty. 
Uncertainties that are entirely quantum do not allow for the definition of sub-classes of e.g.~different momentum values, since there are no hidden variables. It thus has to be concluded that all ultra-relativistic particles of thermal origin have the same momentum of $\langle \hat p_x \rangle \approx 0$.

The product $\,\lambda_{\rm th} \cdot \Delta^{\rm fw}_{\rm th} p_x \approx h$ corresponds to the lower limit of the inequality given by W.~Heisenberg in \cite{Heisenberg1927}. 
The quantum uncertainties of pure states, however, are also described by the lower bound of `Heisenberg's' uncertainty relation according to  Kennard, Weyl and Robertson \cite{Kennard1927,Weyl1927,Robertson1929}. In terms of standard deviations, the lower bound of the latter reads $\Delta \hat x \cdot \Delta \hat p_x = h/(4\pi)$. 
Indeed, the two lower bounds are approximately the same when taking into account, first, that $\,\lambda_{\rm th}$ and $\Delta^{\rm fw}_{\rm th} p_x$ are defined as \emph{full} widths, i.e.~approximately as twice the standard deviations and, second, that the energy distribution within a `wavelength' has two maxima and is roughly a factor of $\pi$ broader than just one of these maxima.
In other words, one can make the rough connections $\Delta \hat x \approx \lambda_{{\rm th}}/(2 \pi)$ and $\Delta \hat p_x \approx \Delta^{\rm fw}_{\rm th} p_x /2$.\\

Below I compare the positional uncertainty of the neutrinos with their average distance at times just before decoupling and find that they are roughly the same. The average neutrino number density per flavor, spin and cubic meter as a function of the temperature of the universe as derived by the (fully classical) Boltzmann equation reads \cite{Sciama1994,Pastor2011}.
\vspace{-2mm}
\begin{equation} \label{eq:n}
n_\nu = 6 \pi \cdot \zeta(3) \left(\frac{k_B T}{h c}\right)^3 \,, 
\end{equation}
\vspace{2mm}
where $\zeta(3) \approx 1.2$. The average one-dimensional distance between two identical neutrinos is thus given by
\vspace{2mm}
\begin{equation} \label{eq:d}
d \approx \sqrt[3]{\frac{1}{6 \pi \cdot \zeta(3)}} \cdot \frac{h c}{k_B T} \approx 0.35 \cdot \frac{h c}{k_B T} \;.
\end{equation}
\vspace{2mm}
At decoupling temperature, one gets $d_{\nu {\rm d}} \approx 1.7 \cdot 10^{-13}$\,m.  
For comparison, the full spread of the neutrino position uncertainty around decoupling was 
\vspace{2mm}
\begin{equation}
\label{eq:2dx}
2 \Delta \hat x_{\nu{\rm d}} \approx \frac{\lambda_{{\rm th,} \nu {\rm d}}}{\pi} = \frac{1}{\pi} \cdot \frac{h c}{k_B T_{\nu {\rm d}}} \approx 0.32 \cdot \frac{h c}{k_B T_{\nu, {\rm d}}} 
\,.
\end{equation}
\vspace{2mm}
At decoupling temperature, one gets $ 2\Delta \hat x_{\nu {\rm d}} \approx 1.5 \cdot 10^{-13} \, {\rm m}$. $ 2\Delta \hat x$ and $d$ turn out to be basically identical, keeping in mind the rough estimations Eq.\,(\ref{eq:2dx}) is based on.\\ [-3mm] 

The result of this subsection is as follows.
During decoupling, when the interactions between individual neutrinos became negligible, the positional uncertainties of identical neutrinos almost overlapped.
This implies that the neutrinos were at the transition to a quantum degenerate Fermi gas. (In the following I will use the term ?field? instead of ?gas?, since the latter could imply a wrong picture of neutrinos localized in space.)
In addition, the eigenvalue spectra of the neutrino positions and impulses were given exclusively by quantum uncertainties of pure states.
It follows that all neutrinos had the same momentum, and the one-dimensional momentum expectation values with respect to the primeval plasma was $\langle \hat p_x \rangle \approx \langle \hat p_y \rangle \approx \langle \hat p_z \rangle \approx 0$. 
Since (local) hidden variables do not exist \cite{Aspect1981}, the conception of neutrinos having a trajectory, i.e.~moving like classical particles, was not adequate at decoupling time. 
Instead, a valid description of neutrinos at and after decoupling has to include interference of position uncertainties and the Pauli exclusion principle.\\

\subsection{The prevailing but incorrect description of neutrinos after decoupling} \label{ssec:dcl}

There is no question that after the decoupling the neutrinos had an undisturbed (`free') evolution in the continuously expanding universe. 
In relevant literature, the evolution of neutrinos is usually described as a free, particle-like propagation of mutually independent neutrinos, which are assigned to different momentum classes. The expansion of the universe has red-shifted the momentum spectrum in the same way as it has red-shifted the frequency spectrum of radiation -- with the following consequence. While at decoupling time, the neutrinos constituted a considerable fraction of the universe's total energy, today the have lost most of their relativistic energy. 
As a matter of fact, the cosmological redshift has maintained the thermal \emph{appearance} of momentum and frequency spectra, although neither neutrinos (nor photons) have had any interaction after decoupling.
For this reason, `temperatures' are conveniently used to describe the relic neutrinos as well as the cosmic micro-wave background, although these `temperatures' do not correspond to actual thermodynamic temperatures but to frozen (and red-shifted) images of the temperatures at decoupling. The `temperature' of today's cosmic neutrino back-ground (C$\nu$B) is derived to $T_{\rm C \nu B} = 1.945$\,K. This value is below the temperature of the CMB of $T_{\rm CMB} = 2.725$\,K, because the electro-magnetic radiation field was reheated when the electrons and positrons of the primeval plasma annihilated shortly after neutrino decoupling and decoupled at 380,000 years later at slightly elevated temperature  \cite{Sciama1994,Pastor2011}.
It is standard in relevant literature to use the `temperature' value of the C$\nu$B and Eq.\,(\ref{eq:n}) to derive the well-known value of today's average relic neutrino number density of  $n_{\nu,0} \approx 56$\,cm$^{-3}$ for each of the six types of neutrinos (two spin values $\times$ three flavors) \cite{Sciama1994,Pastor2011}. Today's average relic neutrino number density is obviously given by the total neutrino number at decoupling and the size of today's universe. The `temperature' in Eq.\,(\ref{eq:n}) simply is a measure of the expansion of the universe without  meeting the requirement of an actual thermodynamic temperature for values below the decoupling temperature.\\ 

The description of the decoupled neutrinos as particles that are assigned to different momentum classes and propagate independently is wrong, because it does not correspond to quantum field physics, according to which a `free evolution? in flat space-time corresponds to a steadily increasing position uncertainty in all three dimensions.
Instead, a valid description of neutrinos during and after decoupling must include overlapping positional uncertainties and interference in accordance with the Pauli principle of exclusion.

The same quantum field physics naturally also applies to all other decoupled or 'weakly interacting' particles such as the hypothetical \emph{weakly-interacting massive particles} (WIMPs) and the \emph{weakly-interacting slim particles} (WISPs) \cite{Goodman1985,Jungman1996,Ringwald2012,Klasen2015,Bertone2018}. 
The prevailing semiclassical, particle-like statistical view of WIMPS and WISPS is also not conform with the fact of quantum field physics that without interactions, position uncertainties become large and therefore interference relevant.

In the next subsections I present my reasoning why the overlapping of neutrino wave functions ultimately led to a much higher average kinetic energy of today's C$\nu$B than the one associated with $T_{\rm C \nu B} = 1.945$\,K. \\

\subsection{After decoupling: The emergence of macroscopic neutrino field-modes}  \label{ssec:qft}

According to quantum physics, as it is mathematically described by quantum field theory, a `neutrino' is the quantized excitation of a Fourier-limited mode of the neutrino field oscillation. The `mode' is understood as a physical object and represented by its wave function. 
In case of fermionic modes, occupation numbers cannot exceed one. 

Before decoupling, the neutrinos belonged to modes that were \emph{distinguishable} due to their different locations in space-time. Energy quanta (neutrinos) got frequently transferred from one mode to another, mainly by annihilation and creation according to Eq.\,(\ref{eq:chain}).
After decoupling, the modes' occupation numbers did not change any more, but the modes became \emph{distinguishable} (`degenerate').  
The free evolution of the modes, which made them degenerate, was fundamentally different from the classical concept of free particle-like propagation. The free evolution of a neutrino mode was governed by momentum uncertainties $\Delta \hat p_x$ (likewise $\Delta \hat p_y$ and $\Delta \hat p_z$) at decoupling, which resulted in a continuous ultra-relativistic increase of the neutrinos' position uncertainty, or in other words, of the spatial mode size: 
\vspace{2mm}
\begin{equation}
\label{eq:xt}
\Delta \hat x_\nu (t) = \Delta \hat y_\nu (t) = \Delta \hat z_\nu (t) \approx \Delta \hat x_{\nu {\rm d}} + c (t - t_{\nu {\rm d}})\,.
\end{equation}
\vspace{2mm}
The consequences were mode overlapping and the emergence of indistinguishability. 
As an example, consider \ emph {two} neutrino modes exactly at the point in time when all energy exchange with the rest of the universe ceased. Due to their interactions in the past they be highly localized to a level of about $10^{-13}$\,m and accidentally have an occupation number of unity. The modes have spherical symmetry with Gaussian energy distribution and are fully identical, except for a spatial separation of $1.7 \cdot 10^{-13}$\,m. Due to the high localization and the low neutrino rest mass, their momentum uncertainty is ultra-relativistic. The radii of the modes' position uncertainties have to grow with almost the speed of light, while their distance hardly changes. After 1\,$\mu$s, the modes have radii of about 300\,m and they fully overlap. Their occupation number in terms of two neutrinos in total does not change.
In the course of overlapping, however, the two modes become indistinguishable, because (i) their separation becomes negligible compared with their position uncertainty, (ii) the expectation values of their momenta remain identical (and small), and (iii) the momentum uncertainties also remain identical. The neutrinos being the mode's excitation quanta are thus also indistinguishable. Their position uncertainty corresponds to the size of the joined (degenerate) mode. 
There is no other possibility how interaction-free physical systems can evolve. 

The question arises as to why this result does not contradict the Pauli principle of exclusion. The answer is given by the internal structure of the new highly degenerate mode, which is a consequence of interference. The two neutrinos are excitation quanta of the \emph{same} (new) mode, and as such indeed indistinguishable. The internal structure of the mode, however, is such that it guarantees that the two neutrinos would always localize with a spatial separation. 
Here, I use the word 'would' to point out the fact that localization requires an interactions, which, as requested for this example, has zero probability. 
Uncertainties with internal correlations (here: anti-correlations) represent `quantum correlations', i.e~entanglement \cite{Schnabel2020} (see also subsection \ref{ssec:pairing}). A very important insight is the fact that quantum anti-correlation does not demolish indistinguishability, since the individual properties of the fermions remain undetermined.

It has to be concluded that one microsecond after neutrino decoupling, the universe was filled with a huge number of spherical, mutually overlapping degenerate neutrino fields of $\Delta \hat x {\rm (1}\mu{\rm s)} = 300$\,m radii. Due to particle number conservation, each was occupied by up to $N \approx 10^{46}$ indistinguishable neutrinos. As in the example above, each mode evolved without violating the Pauli exclusion principle. The emerged internal quantum anti-correlation had to have the property of a three-dimensional standing wave, since it was produced from spherical waves from all directions. The high neutrino density enforced a wavelength that was (almost) as short as the thermal De\,Broglie wavelength at decoupling time ($\lambda = \lambda_{{\rm th},\nu d}$). In the nodes, the probability of finding a neutrino was zero. The locations of the nodes, however, were not determined with respect to the primeval plasma. As in the example above, these quantum correlations guaranteed spatial separation of the neutrinos in case they got localized due to the unlikely case of interactions.

While the neutrino uncertainties grew, inhomogeneities of the neutrino energy density on scales smaller than the neutrino fields got smeared out. However, inhomogeneities on larger scales persisted for a while. Spacetime was not completely flat, especially due to the mass contribution of the neutrino fields. Obviously, neutrino modes experienced the attractive potentials due to their own masses. 
Since all field modes continuously evolved further to larger sizes and larger neutrino numbers $N$, while the universe was continuously expanding, it seems plausible that the neutrino fields began to localize due to the space-time curvature they created themselves.

\subsection{Gravitational self-localization and temperature of the neutrino field-modes}  

By `self-localization' I do not mean a gravitational collapse of the neutrino field, but the establishment of a homogeneous  energy density that is stably positioned around its own center of mass.  A `collapse' would have increased the density of the neutrino field, which was not possible without violating the Pauli exclusion principle. The crucial question in this scenario is whether the thermodynamic temperature of the neutrino field after decoupling $T_\nu (t \!>\! t_{\nu {\rm d}})$ was low enough for gravitational self-localization. My answer to this question is worked out here, starting with the neutrino temperature.

When the neutrinos were in thermal equilibrium with the universe, the joint temperature was connected to the age of the universe $t$ in the following way \cite{Sciama1994}
\vspace{2mm}
\begin{equation}
\label{eq:t}
T / {\rm K} = \frac{2 \cdot 10^{10}}{\sqrt{t / {\rm s}}}  \,.
\end{equation}
\vspace{0mm}
The temperature ($T = T_\nu$) was directly linked to the weak-force scattering (collision) rate per neutrino \cite{Sciama1994}
\vspace{2mm}
\begin{equation}
\label{eq:f}
f_{\nu,{\rm s}} \!=\! (T_\nu /T_{\nu {\rm d}})^5 / t_{\nu {\rm d}} \,.
\end{equation}
\vspace{0mm}
At decoupling time  $t_{\nu {\rm d}} = 0.5$\,s,  $f_{\nu,{\rm s}}$ was as small as 2\,Hz. This fact does correspond to `decoupling', since the time span of a single collision started exceeding the age of the universe.

In equilibrium, every degree of freedom is excited by the same energy, 
which is $k_B T$ for ultra-relativistic particles.
Eq.\,(\ref{eq:f}) allows to calculate the quantity `average action per neutrino' $S_{\nu,{\rm s}} (T_\nu)$.
\vspace{2mm}
\begin{equation}
\label{eq:L}
\hspace{-7mm} S_{\nu,{\rm s}} (T_\nu) =  \frac{k_B T_{\nu}}{f_{\nu,{\rm s}}} = \frac{k_B T^5_{\nu {\rm d}}} {T_{\nu}^4} t_{\nu {\rm d}} \; \hspace{3mm} {\rm with}\,\, S_{\nu,{\rm s}} \ge h
\,.
\end{equation}
\vspace{2mm}
The existence of a lower bound is given by the fact that there is a smallest possible action -- Planck's quantum of action $h$ \cite{Planck1900}. 
Remarkably, the lower bound of Eq.\,(\ref{eq:L}) sets a relation between neutrino decoupling temperature $T_{\nu {\rm d}}$ and an \emph{upper} temperature bound, beyond which neutrino weak-force collisions are unphysical, since $S_{\nu,{\rm s}} < h$.
Taking the value $T_{\nu {\rm d}} = 3 \cdot 10^{10}$\,K, the highest possible weak-force collision rate is reached at $T_{\rm ew} \approx 4 \cdot 10^{15}$\,K. 
This value corresponds to the well-known electroweak scale of around 246\,GeV ($ \approx 3\cdot 10^{15}$\,K). This energy scale is typical for processes described by the electroweak theory, which unifies the weak force and the electromagnetic force to form the electroweak force.\\

An estimation for $T_\nu(t)$ at times \emph{after} decoupling is not obvious. 
The sections above, however, give several hints that the thermodynamic temperature of ultra-relativistic particles is linked to minimal quantum uncertainties.
The temperature of neutrinos is bounded from above by $h$, which is linked to the minimal uncertainty product. And during decoupling the thermal De\,Broglie wavelength turns out to be equivalent to a pure-state position uncertainty $\Delta \hat x$. 
Using Eq.\,(\ref{eq:lambda}), the temperature of an ultra-relativistic particle field at decoupling can be expressed by
\begin{equation}
\label{eq:T1}
\hspace{-10mm}
T (t) \approx \frac{hc}{k_B \lambda_{{\rm th}}(t)}  \approx \frac{hc}{k_B 2 \pi \Delta \hat x (t)} =  \frac{\hbar c}{k_B \Delta \hat x (t)}  \; ,
\end{equation}
where $\hbar = h/(2 \pi)$ is the reduced Planck constant.
I don't see any reason why this relationship should no longer be valid after decoupling.
After decoupling, the neutrino position uncertainty evolved according to Eq.\,(\ref{eq:xt}). 
I thus propose the following expression for the temperature of the neutrino field after decoupling
\vspace{2mm}
\begin{equation}
\label{eq:T}
\hspace{-10mm}
T_{\nu}(t) 	=  \frac{\hbar c}{ k_B (\Delta x_{\nu {\rm d}} \!+\! c(t - t_{\nu {\rm d}} )) } 
	        \approx \frac{\hbar}{k_B (t - t_{\nu {\rm d}} ) } \; , \;\;{\rm for} \hspace{1.8mm} t- t_{\nu {\rm d}} \gg  \frac{\Delta x_{\nu {\rm d}}}{c}   \,.
\end{equation}
\vspace{2mm} 
Accordingly, the neutrino field is much colder than the `temperature' assigned to the cosmic neutrino background ($T_{\rm C \nu B} = 1.945$\,K,), which is not surprising, since the latter does not take into account the greatly reduced collision rate.
As an example, 0.4\,s after decoupling the neutrino field temperature was $T_{\nu}(0.9\,{\rm s}) \approx 0.02$\,nK, and accordingly $k_B T_{\nu}(0.9\,{\rm s}) \approx 1.6\cdot 10^{-15}$\,eV. \\


%
Other systems that are decoupled from the rest of the universe are black holes. 
Having mentioned this, Eq.\,(\ref{eq:T}) provides a remarkable result:
An ultra-relativistic neutrino field with uncertainty `radius' $\Delta \hat x_\nu \approx c(t - t_{\nu {\rm d}})$ has the same temperature as a black hole with Schwarzschild radius $r_{\rm S} = \Delta \hat x_\nu/(4 \pi)$: 
\vspace{2mm}
\begin{equation}
\label{eq:HT} \hspace{-20mm}
T_{\nu}(\Delta \hat x_\nu) 	\approx  \frac{\hbar c}{\Delta \hat x_\nu k_B}  \;=\;   T_{\rm BH}(r_{\rm S}) =  \frac{\hbar c^3}{8 \pi G M(r_{\rm S}) k_B} =  \frac{\hbar c}{4 \pi r_{\rm S} k_B}  \;\;\;{\rm for} \;\;\;\Delta \hat x_\nu = 4 \pi r_{\rm S} \, , 
\end{equation}
\vspace{2mm} 
where $T_{\rm BH}(r_{\rm S})$ corresponds to the Hawking temperature \cite{Hawking1974}, $G$ is the gravitational constant, $M$ the mass of the black hole, and $r_{\rm S} = 2GM/c^2$ \cite{Schwarzschild1916b}.
The temperatures are of the same order of magnitude, which supports my suggestion for the neutrino field temperature in Eq.\,(\ref{eq:T}). 
Note that $\Delta \hat x_\nu$ and $r_{\rm S}$ should not be identical, because $r_{\rm S}$ is proportional to the mass of the black hole and corresponds to its radius as defined from the \emph{outside}.\\

Having found an approximation for the thermodynamic temperature of the neutrino field after decoupling (Eq.\,(\ref{eq:T})) it is possible to estimate whether gravitational self-localization of neutrino fields was possible. Below, I show that this was inevitable.
Self-localization of the neutrinos had to take place without violating Pauli's principle of exclusion. The interference of the positional uncertainties of the neutrinos after the decoupling had already taken this into account. The boundary condition for the gravitational self-localization of the large, highly excited neutrino fields was an unchanged particle density. 
In this scenario, the gravitational self-localization can be approximated by the known statistics of indistinguishable, non-interacting boson particles:\\
For simplification, I consider just two energy levels and a total number of $N$ neutrinos. The lower level corresponds to the quanta's ground state in the envisioned trapping potential of gravitational self-localization. The second level corresponds to an elevated neutrino energy of flat space-time. The potential difference be $|E|$.
The probability of having an upper level population of $K < N/2$ with $N \rightarrow \infty$ is given by $P(K) \approx  e^{-K |E| / (k_BT_\nu)}$. 
It can be shown that the upper level population itself is given by $K \approx e^{- |E| / (k_BT_\nu)}/(1-e^{- |E| / (k_BT_\nu)})$, which is independent of $N$. 
For every small value of $|E| > 0$, $K$ becomes negligible as $N$ approaches infinity.
In practice, neutrino field modes reached extremely large, but nevertheless \emph{finite} values of $N$. For this reason, the gravitational trapping potential must $|E|$ be quantified.

My quantitative estimation of the trapping potential $|E|$ refers to the potential of a black hole. In section \ref{sec:smBH}, I show that 0.9\,s after the Big Bang the first primordial black holes emerged. In this section I am referring to the fact that galaxies have not only dark matter halos but also supermassive black holes at their centres. For a massive neutrino field that accumulates around a black hole, the trapping potential is $|E| = mc^2$, with $m$ the neutrino rest mass.
In fact, there are three different rest masses of neutrinos.
Whereas the three neutrino \emph{flavours} are the eigenvalues of the weak interaction, the three neutrino \emph{rest masses} are eigenvalues of neutrino propagation. 
The observation of neutrino oscillations showed that all these masses are non-zero. The individual values, however, are not known, just sums and differences of their squares \cite{Cuesta2016,Capozzi2016}.  \\ 
Using the lower bound for the average neutrino rest mass of $m \approx 0.033\,{\rm eV}/c^2$ 
as derived from the observation of neutrino oscillations \cite{Cuesta2016}, $|E|$ is thirteen orders of magnitude greater than $k_B T_\nu (0.9\,{\rm s})$. 
In this case, $K$ is zero and not a single neutrino is not localized around the black hole.
The lowest neutrino mass, however, might be much lower than the average neutrino mass. 
In Sec.\,\ref{sec:mass}, I present a very conservative estimation of the value for the lowest neutrino mass and find a lower bound of about $10^{-26}$\,eV/$c^2$. Consequently, $|E|$ is at most eleven orders of magnitude lower than $k_B T_\nu (0.9\,{\rm s})$. 
In this case $K \approx 10^{11}$.
To conclude, even in case of an extremely low neutrino rest mass, $K$ is negligible compared to the typical numbers of $N$. \\

In summary, only a fraction of a second after decoupling were essential parts of the macroscopic neutrino fields in their quantum mechanical ground states, localized and trapped by their own mass. In the case of fermionic fields, such states are referred to as `Fermi seas'.
The near zero temperature along with the fermionic nature of the field forced density fluctuations of almost zero across the dimensions of the Fermi sea.
Even quantum fluctuations in density were suppressed.
Zero density fluctuations imply zero compressibility and superfluidity of the mass.

\subsection{Neutrino pairing} \label{ssec:pairing} 

A Fermi sea is unstable to pairing of its quanta if there is a weak force of attraction between them.
This is the assertion of the Bardeen, Cooper and Schrieffer (BCS) theory \cite{Bardeen1957}, which was formulated to describe the formation of Cooper pairs from initially distinguishable electrons \cite{Cooper1956} and superconductivity in metals at low temperatures.
My proposal is that the neutrinos in Fermi seas are also paired because of the tiny gravitational attraction between them.

The physical description of pairing of initially independent quanta starts from three preconditions. First, the quanta are indistinguishable, from which follows that decoherence is negligible. Second, there exists a mutually attractive force. And third, the first two preconditions last for an expanded period of time in order to take into account the spatial spread of the quanta as well as to counterbalance the finite nature of the attractive force. All three preconditions are fulfilled for a neutrino Fermi sea. 
Indistinguishability guarantees that all quanta have maximal spatial overlap of their position uncertainties, simply because they excite the same spatial mode. Maximal spatial overlap is relevant since it maximises the efficacy of the attractive force. 
The attractive force itself is relevant because it produces an attractive potential.

The formation of pairs of particle is only manifested in case of interactions with the environment that lead to localization.
Neutrinos of a Fermi sea do not have such interactions, however, we can consider such interactions hypothetically.
Since we postulate indistinguishability, we have to conclude the following.
If an interaction localizes a neutrino `$A$' precisely at position $\vec{x}_A$ then there is another neutrino `$B$' at precisely the same position with certainty, i.e.~100\% probability (regardless whether the second neutrino is separately detected or not). 
This is enforced because of the vanishing quantum uncertainty in the differential position $\vec{x}_A - \vec{x}_B = 0$ and the attractive force.
If the same interaction additionally determines the spin value, the second neutrino has opposite spin, again with certainty. This is the anti-correlation enforced by the Pauli principle, which leads to bosons of zero spin.
Alternatively, the interaction may determine a neutrino with a precise momentum value $\vec{p}_A$ with respect to the centre of mass of the neutrino field. In this case, the interaction must realize a second neutrino that has precisely the same momentum but with opposite sign. This is due to indistinguishability combined with momentum conservation. In the concept of pairing, the interaction with the environment affects exactly two neutrinos while all others remain unaffected and thus indistinguishable. The total momentum of all remaining $N-2$ indistinguishable neutrinos has to remain zero, as it was the total momentum of $N$ neutrinos before the interaction. In conclusion, there is no uncertainty in the sum of the momenta $\vec{p}_A + \vec{p}_B = 0$. A simultaneous spin measurement would again show opposite spins. The zero-spin bosons thus have precisely zero momentum. 
Using the concept of pairing provides a complementary view on a neutrino Fermi sea, which is that of a Bose-Einstein condensate of zero-spin particles.

Finally, I consider a large number of such neutrino measurements with perfect quantum efficiency. 
The measurements correspond to an ensemble measurement of identical pairs.
Regardless whether $\vec{x}_A$ or $\vec{p}_A$ is measured, it allows to predict the corresponding value for the corresponding second neutrino without any uncertainty. This is correctly described by quantum theory, since the relevant quantities commute ($[x_A - x_B, p_{x,A} + p_{x,B}] = 0$ for all three dimensions). The measurements demonstrate Einstein-Podolsky-Rosen (EPR) entanglement \cite{Einstein1935}. The physics that describes the evolution to pairs from initially individual systems thus describes the emergence of EPR entanglement. 
(For comparison I refer to \cite{Schnabel2015}, which investigates the physics of the emergence of EPR entanglement between two massive mirrors with respect to their motional degree of freedom in one dimension.)

\subsection{Dark matter mass and the total mass of the neutrino-pair quantum fields} 
\label{ssec:dmm}

A well-known problem with neutrinos as dark matter is that their total rest mass in the visible universe is much less than the mass of dark matter.
The prevailing opinion is that also the sum of rest mass and today's relativistic mass is too small to make up a substantial part of the DM, since the neutrinos are said to have lost most of their relativistic mass through the expansion of the universe.
However, given the gravitational self-attraction of neutrino fields, this assumption should be wrong.
My argumentation is as follows.
First, dense neutrino fields, as they spread over sizes as large as galaxy clusters and beyond, resembled a strong gravitational force that locally acted against the expansion of the universe.
Second, there were also space-time volumes containing less neutrino field energy. At these space-time locations the dilution of energy density was faster due to higher expansion rate. In other words, the initial quantum fluctuations in the density got amplified and the expansion rate of the universe became more and more inhomogeneous as time progressed. From this point of view, the well-known cosmic voids \cite{Geller1989} seem to be a natural consequence. 

While electromagnetic radiation experienced a redshift due to the average expansion of the universe, the neutrino fields were predominantly in space-time regions of reduced expansion. They experienced a reduced redshift, which was given by the average of the local expansions, weighted with the neutrino field density distribution.
In the following I estimate the average rest-mass density of the decoupled neutrinos and from that the cumulated factor by which the expansion of the universe was reduced on average at locations of dense neutrino fields in order to fit to the total dark matter mass that we observe today.  
Today's average neutrino/anti-neutrino number density amounts to $3.4 \cdot 10^{8}$\,m$^{-3}$, taking into account two spin and three mass values \cite{Sciama1994,Pastor2011}, see also Eq.\,(\ref{eq:n}) for $T = T_{\rm C \nu B} = 1.945$\,K.
Using the upper bound for the average neutrino rest mass of $0.04\,{\rm eV}/c^2$ ($\approx 7 \cdot 10^{-38}$\,kg) \cite{Palanque-Delabrouille2015} the average rest-mass density is less than $2.4 \cdot 10^{-29}$\,kg/m$^3$.
Using the lower bound for the average neutrino rest mass of $0.033\,{\rm eV}/c^2$ ($\approx 6 \cdot 10^{-38}$\,kg) as derived from the observation of neutrino oscillations \cite{Cuesta2016}, the total rest mass of all neutrinos is at least $2 \cdot 10^{-29}$\,kg/m$^3$. 
The average DM mass density, however, is more than a hundred times higher, namely about $2.3 \cdot 10^{-27}$\,kg/m$^3$. This value is given by the data shown in Fig.~\ref{fig2} and the total energy density of the universe. Due to the \emph{WMAP} mission \cite{Komatsu2011} it is known that the latter almost corresponds to the (mass equivalent) critical density of a flat universe of $\rho_{\rm crit} \approx 9 \cdot 10^{-27}$\,kg/m$^3$.

If the decoupled neutrinos constitute all cold dark matter, their relativistic energy today needs to be about a hundred times higher than their average rest mass. 
In other words, the effective expansion of the `neutrino-field universe' must have reduced the average relativistic energy in three dimensions from initially $3 \times 2.6$\,MeV to now about $3 \times 1.3$\,eV ($\approx 100 \times 0.04$\,eV), which corresponds to an average neutrino field red-shift of $z^\nu_{\nu {\rm d}} = 2.6\,{\rm MeV} / 1.3\,{\rm eV} \approx 2 \cdot 10^{6}$ instead of the literature value of $\tilde{z}^\nu_{\nu {\rm d}} =  T_{\nu,{\rm d}} / T_{\rm C \nu B}  \approx 1.5 \cdot 10^{10}$. 

The summary of my hypothesis about the nature and origin of cold dark matter is as follows.
Cold DM as observed today are neutrino fields with a temperature extremely close to zero. Its total mass is dominated by the relativistic mass.
The quantised excitation of the neutrino field can be understood as (degenerate) neutrino pairs.
Neutrino pairs are not strictly localised to a single galactic halo but are excitations of connected parts that form a DM scaffold.

\subsection{Quantum field view on neutrino decoupling} \label{ssec:decoupling}

\emph{The neutrinos decoupled when the average weak-force collision rate of the individual neutrinos dropped below the expansion rate of the universe.} 
But what happened to the neutrino-pair creation and annihilation rates?
The thermal energies of electrons and positrons remained high enough for neutrino pair production. 
But a continued neutrino pair production would have resulted in a huge energy loss of the Universe since all these neutrinos would have decoupled as well. The universe would have cooled quickly to a thermal energy of about 0.5\,MeV ($6 \cdot 10^9$\,K), at which the thermal production of electron/positron pairs had to stop and all existing such particles recombined.
This indeed happened, but only due to the continued expansion of the universe and only at $t \approx 11$\,s \cite{Sciama1994}. 

This subsection complements the physics of neutrino decoupling by the quantum field view, which explains why the neutrino production stopped basically at $t_{\nu {\rm d}}$ although the temperature of the universe was still high enough for neutrino-pair production at high rate.

As outlined in subsection \ref{ssec:qft}, neutrinos are the excitation quanta of neutrino field modes. At any temperature $T$ of the early universe, these modes expanded with almost the speed of light due to the ultra-relativistic momentum uncertainty. 
The only way of avoiding the overlap of excited modes, which would have violated Pauli's exclusion principle, was a high rate of their depopulation.
The required rate can be estimated. According to Eq.\,(\ref{eq:d}), relevant overlapping would have occurred already after a time interval of about $(3 \cdot 10^{-22} T_{\nu {\rm d}}/T)$\,s. During this period of time, one out of two populated modes had to be depopulated, which required a depopulation rate of slightly more than $(10^{21} T/T_{\nu {\rm d}})$\,Hz per neutrino. The weak force collision rate of individual neutrinos of $T^5/(T^5_{\nu {\rm d}}  t_{\nu {\rm d}})$ was too low.
The highest possible interaction frequency of a relativistic gas in thermal equilibrium is generally given by $k_B T /h$. This value can be written in relation to the neutrino decoupling temperature as 
\vspace{2mm}
\begin{equation}
\label{eq:G}
\Gamma \approx 
\frac{k_B T}{h} \approx 6 \cdot 10^{20} \frac{T}{T_{\nu {\rm d}}}\; {\rm Hz} \; , \;\;  {\rm for} \; T \ge T_{\nu {\rm d}} \,.
\end{equation}
\vspace{2mm}
Since the weak-force collision rate at decoupling was much lower, the above equation must be related to the rate of neutrino-pair creation and annihilation. 
The population and depopulation rates of \emph{individual} neutrino modes are twice as high, which indeed meets the value of the depopulation rate requested above.
My conclusion requires that neutrino-pair creation and annihilation at ultra-relativistic temperatures corresponds to an `action' as small as $h$.
This finding is justified since Eq.\,(\ref{eq:chain}) describes \emph{coexisting} forms of energy, all degrees of freedom are excited by $k_B T$, and the tiny quant of action `$h$' makes the difference between a neutrino pair and an electron/positron pair.

The question of why the neutrino-pair creation and annihilation stopped seems to be even more difficult to answer in view of the high rate in Eq.\,(\ref{eq:G}).
Also needs the question to be answered what mechanism balanced creation and annihilation in thermal equilibrium.
It is easy to imagine that the annihilation of electron-positron pairs and thus the creation of neutrino pairs happened at a high rate since the electro-magnetic attraction of electrons and positrons supported their annihilation. 
It is less obvious, what mechanism resulted in an equally high rate for neutrino-pair annihilation. 
My answer to that question is the following.
The parity of Eqs.\,(\ref{eq:d}) and (\ref{eq:2dx}) shows that the neutrino position/momentum phase space before decoupling was `almost' packed.
Here, I slightly refine this statement.
In fact, balancing between creation and annihilation, requested 50\% of all modes that were addressed by the neutrino pair creation were \emph{not} occupied.
In this situation, 50\% of all neutrino-pair creation processes indeed resulted in the population of field modes, but in 50\% of the neutrino pair `creation' processes an occupied modes were addressed and the Pauli exclusion principle enforced destructive interference. In other words, in 50\% of the creation processes, the net effect was neutrino annihilation. 
This way, creation and annihilation were balanced even when the weak-force collision rate of individual neutrinos was already small. 

Finally, I answer the question why the huge rate of neutrino-pair creation stopped at decoupling although the temperature of the universe was still high enough for neutrino-pair production at high rate.
At decoupling, the contribution to the depopulation of neutrino modes by the weak-force scattering became that low, that the expansion of populated neutrino modes occupied the entire space-time. New neutrino pairs could not be created by the annihilation of electron/positron pairs because of the Pauli exclusion principle.
I thus propose the following complementary description of neutrino decoupling. \emph{The neutrino decoupling temperature $T_{\nu {\rm d}}$ is the one, at which the depopulation rate of occupied neutrino modes dropped below their relative expansion rate.}\\

\begin{figure}
	\centering
	\vspace{2mm}
	\includegraphics[width=0.62\textwidth]{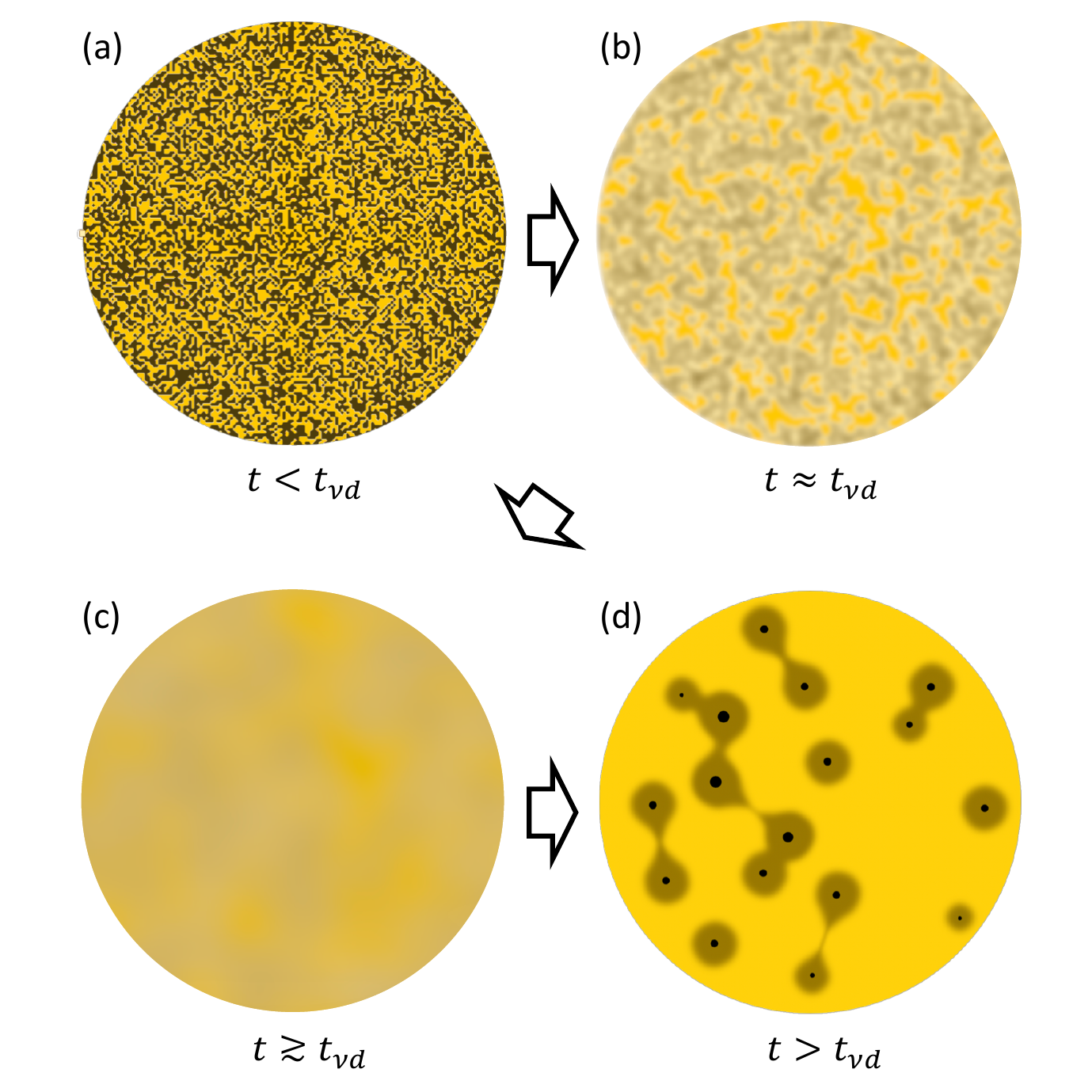}
	\vspace{0mm}
	\caption{Hypothesis of neutrino-to-dark-matter evolution starting with neutrino de\-coup\-ling at $t_{\nu {\rm d}} \approx 0.5$\,s after the Big Bang. 
	(a) Highly localized neutrinos (black) that are strongly coupled to the rest of the universe via the annihilation of electron/positron pairs.
	(b) Already delocalized neutrino matter (grey shades) shortly after instantaneous decoupling.
	(c) Largely delocalized neutrino-field Fermi seas. 
	(d) Dark-matter halos around primordial supermassive black holes. The latter are due to a direct transformation of the Fermi seas. Gravitational collapses were not necessary. Overlapping DM halos provide a gravito-covalent bond, which has been relevant for galaxy cluster formation and stability.}
\label{fig3}
\end{figure}

\section{Primordial formation of super-massive black holes} \label{sec:smBH}
The direct and necessary consequence of large-scale neutrino Fermi seas was the primordial formation of super-massive black holes. 
Their formation did not require any agglomerations of mass via the gravitational force, i.e.~no gravitational collapses, since the mass/energy density of the neutrino field was already sufficient for the emergence of event horizons.  
As a matter of fact, the mass/energy density in the early universe, at any time, was that high that the integration over finite volumes resulted in values beyond the critical value for the emergence of event horizons. In this section, I present my argumentation why black holes did rarely emerge before neutrino decoupling, but necessarily emerged right after neutrino decoupling.

The well-known necessary condition for the emergence of a black hole can be formulated as follows. 
For a black hole without spin and charge (a `Schwarzschild black hole'), the total mass within a volume of radius $r_{\rm S}$ (the `Schwarzschild radius') has to be above the threshold 
\vspace{2mm}
\begin{equation}
\label{eq:BH}
M_{\rm BH} =  r_{{\rm S},{\rm BH}}\, c^2/(2G) \approx r_{{\rm S},{\rm BH}} M_\odot / r_{{\rm S},\odot} \, . 
\end{equation} 
\vspace{2mm}
where $G$ is the gravitational constant, $M_\odot \approx 2\! \cdot \!10^{30}$\,kg is the mass of our sun and $r_{{\rm S},\odot} \approx 3\,{\rm km}$ its Schwarzschild radius.

The question arises, why the condition above is necessary but not \emph{sufficient} for BH formation.
My answer is the following. The energy in finite volumes of the early universe fluctuated by too much and too quickly. The continuous redistribution of mass/energy density in the early universe frustrated the necessary condition for BH formation to remain fulfilled for a sufficiently long time.  
The emergence of event horizons competed with the emergence of those of partially overlapping volumes. 
An important issue was the finite speed of light. A direct consequence of Einstein's special theory of relativity \cite{Einstein1905srt} is the fact that any kind of information can maximally travel at the speed of light, because otherwise the principle of cause and effect (`causality') would be violated. 
The above condition becomes only necessary \emph{and} sufficient, if it persists for the same unambiguously defined volume over a time period that light requires to propagate over the distance that corresponds to the Schwarzschild radius. 
In the early universe, the sufficient condition for BH formation was not met. (It cannot be completely ruled out that in all phases of the early universe a small number of black holes "accidentally" formed, so to speak.)  
But the situation changed completely, once the neutrinos had decoupled. Neutrino fields of large dimension had a well-defined centre of mass location in spacetime and showed only small and slow internal energy redistributions; neutrino fields in their ground states did not show \emph{any} energy redistributions, not even due to quantum fluctuations.

In the following, I quantitatively consider the sufficient condition of the primordial formation of BHs and find a lower bound to their mass.
Required is the expression for the total neutrino mass density at and after decoupling time when the neutrinos were ultra-relativistic. 
Neglecting the neutrino rest mass, Eq.\,(\ref{eq:n}) yields for all six neutrino species and all three degrees of freedom at decoupling time $t_{\nu {\rm d}}$
\vspace{2mm}
\begin{equation}
\label{eq:Omega}
\hspace{0mm}
\Omega_\nu (t_{\nu {\rm d}})  \approx 6 n_{\nu {\rm d}} \frac{\langle |\vec{p}\,| \rangle_{\nu {\rm d}}}{c} \approx 108 \pi \cdot \zeta(3) \frac{(k_B T_{\nu {\rm d}})^4}{h^3 c^5} 
\approx  1.7 \cdot 10^{10}\,\frac{{\rm kg}}{{\rm m}^3} 
\,. 
\end{equation} 
\vspace{2mm}

After decoupling, the above mass density got reduced due to the expansion of the universe. The one-dimensional expansion of the universe 
is given by the scale factor $a = 1/(1+z)$. Generally, relativistic energy densities are inversely proportional to the increase in the volume and to the red shift $z$, i.e.~they are proportional to $a^{-4}$. A large period of time around neutrino decoupling is known as the radiation-dominated era, where $a(t) \propto t^{1/2}$, with $t$ again being the age of the universe \cite{Sciama1994}. 
The average mass-equivalent relativistic neutrino energy density at least for several seconds after decoupling can thus be approximated to 
\vspace{2mm}
\begin{equation}
\label{eq:Omegat}
\hspace{0mm}
\Omega_\nu (t)  \approx 108 \pi \cdot \zeta(3) \frac{(k_B T_{\nu {\rm d}})^4}{h^3 c^5}  \left ( \frac{t_{\nu {\rm d}}}{t} \right )^{\!\!2}
\approx  1.7 \cdot 10^{10}\, \left ( \frac{t_{\nu {\rm d}}}{t} \right )^{\!\!2} \,\frac{{\rm kg}}{{\rm m}^3} 
\,. 
\end{equation} 
\vspace{2mm}
What needs to be added is the average total mass equivalent energy density of the rest of the universe, the primeval plasma without neutrinos, since it also contributed to the formation of black holes. The corresponding $\,\Omega_{pp}\,$ was mainly given by the degrees of freedom of the other relativistic species, i.e.~electrons, positrons and photons. Its time dependence is thus equivalent to the one of $\Omega_\nu (t)$, at least for the relevant few seconds after decoupling.
At any time, the total energy density of the Universe has been large enough for the formation of black holes when integrated over a sufficiently large sphere. Eq.\,(\ref{eq:Omegat}) yields
\vspace{2mm}
\begin{equation*}
\label{eq:rs0}
\hspace{0mm}
(\Omega_\nu (t) + \Omega_{pp} (t)) \cdot \frac{3}{4} r^3_{{\rm S},{\rm BH}} (t) = r_{{\rm S},{\rm BH}}(t) \frac{M_\odot}{r_{{\rm S},\odot}} \; .
\end{equation*} 
\vspace{2mm}
Since the energy densities are proportional to $1/t^2$, the necessary condition for black hole formation at time $t$ was fulfilled within the radius
\vspace{2mm}
\begin{equation}
\label{eq:rs1}
\hspace{0mm}
r_{{\rm S},{\rm BH}}(t) = \sqrt{ \frac{3 M_\odot}{4\pi r_{{\rm S},\odot} \cdot (\Omega_\nu (t_{\nu {\rm d}}) + \Omega_{pp} (t_{\nu {\rm d}}) )} } \cdot   \frac{t}{t_{\nu {\rm d}}}\,. 
\end{equation} 
\vspace{2mm}
The above equation is \emph{sufficient} for black hole formation if energy density fluctuations only exist over distances larger than $r_{{\rm S},{\rm BH}}(t)$ and thus over time periods larger than $r_{{\rm S},{\rm BH}}(t) / c$.
Since neutrino field Fermi seas represent energy densities without any fluctuations, the formation of primordial black holes became likely when the radii of Fermi seas $\Delta \hat x_\nu (t) = c (t - t_{\nu {\rm d}})$ became as large as $r_{{\rm S},{\rm BH}}(t)$. 
The time when the first primordial black holes formed is thus given by
\vspace{2mm}
\begin{equation}
\label{eq:tBH}
\hspace{0mm}
t_{\rm pBH}^{\rm min} = \frac{c t^2_{\nu {\rm d}}}{c t_{\nu {\rm d}} - \sqrt{ 3 M_\odot / \left [4\pi r_{{\rm S},\odot} \cdot (\Omega_\nu (t_{\nu {\rm d}}) + \Omega_{pp} (t_{\nu {\rm d}})) \right ] } } \approx 0.9\,{\rm s}
\; ,
\end{equation} 
\vspace{2mm}
where I use the approximation $\,\Omega_{pp} (t_{\nu {\rm d}}) \approx \Omega_\nu (t_{\nu {\rm d}})$. The first primordial black holes formed about 0.4\,s after neutrino decoupling. Inserting Eq.\,(\ref{eq:tBH}) into Eq.\,(\ref{eq:rs1}) yields the corresponding Schwarzschild radius
\vspace{2mm}
\begin{equation}
\label{eq:BHr}
r_{{\rm S, pBH}}^{\rm min} = \left( \sqrt{\frac{4 \pi \, r_{{\rm S},\odot} (\Omega_{pp} + \Omega_{\nu {\rm d}})}{3 M_\odot}} - \frac{1}{c\, t_{\nu {\rm d}}} \right)^{-1}
\approx \, 1.2 \cdot 10^8 \,{\rm m} \, ,
\end{equation} 
\vspace{2mm}
and the corresponding black hole mass
\vspace{2mm}
\begin{equation}
\label{eq:BHm}
M_{{\rm pBH}}^{\rm min} \approx M_\odot \cdot r_{{\rm S, pBH}}^{\rm min} / r_{{\rm S},\odot} \, \approx  \, 4\cdot 10^4 M_\odot \, ,
\end{equation} 
\vspace{4mm}
again with the approximation $\,\Omega_{pp} (t_{\nu {\rm d}}) \approx \Omega_\nu (t_{\nu {\rm d}})$. $4 \cdot 10^4$ solar masses correspond to a small but never the less \emph{supermassive} black hole, whose origin has been unknown so far.
Eqs.\,(\ref{eq:BHr}) to (\ref{eq:BHm}) are lower bounds. Smaller primordial black holes at earlier times could not form. This result fits in an excellent way to the mass of the smallest currently known supermassive black hole of  $5 \cdot 10^4 M_\odot$ \cite{Baldassare2015}.

Larger black holes than those in Eqs.\,(\ref{eq:BHr}) to (\ref{eq:BHm}) could be formed at later times. The reason is the afore discussed self-amplifying inhomogeneity of the expansion of the universe. At decoupling, energy density fluctuations had purely thermal origin. 
After decoupling, regions with low neutrino energy densities expanded faster, which further diluted the neutrino energy density in these regions and so on. In volumes of diluted neutrino energy density, Fermi seas emerged at later times resulting in larger Schwarzschild radii and correspondingly larger primordial black-hole masses.\\

I now consider two well-known supermassive black holes, which both are above the lower bound.
The supermassive black hole in the centre of the Milky Way has a mass of $4 \cdot 10^6 M_\odot$. Its Schwarzschild radius is $1.2 \cdot 10^{10}$\,m. 
From the perspective of flat space time, the light travel time over the radius is 40\,s, which means that a black hole of this size formed definitely later than 40\,s after decoupling. If one deliberately assumes a formation at $t_{\rm MW} = 88$\,s, the relativistic energy density since decoupling got reduced by $(88/0.5)^2$.
Eq.\,(\ref{eq:Omegat}) together with the approximation $\,\Omega_{pp} (t_{\nu {\rm d}}) \approx \Omega_\nu (t_{\nu {\rm d}})$ yields an energy density of about $10^6\,{\rm kg/m}^3$. Integrated over a sphere with a radius of $1.2 \cdot 10^{10}$\,m yields the consistent value for the black hole mass of $4 \cdot 10^6 M_\odot$.  

An even larger supermassive black hole is located in our neighboring galaxy M87 \cite{Akiyama2019}. It has a mass of $6.5 \cdot 10^9 M_\odot$ and its Schwarzschild radius is $2 \cdot 10^{13}$\,m, which corresponds to a light travel time of 18.5\,h. 
If one deliberately assumes a formation after $t_{\rm M87} = 41$\,h, the relativistic energy density got reduced by $8.7 \cdot 10^{10}$.
Eq.\,(\ref{eq:Omegat}) together with the approximation $\,\Omega_{pp} (t_{\nu {\rm d}}) \approx \Omega_\nu (t_{\nu {\rm d}})$ yields an energy density of about $0.4\,{\rm kg/m}^3$. Integrated over a sphere with a radius of $2 \cdot 10^{13}$\,m yields a mass of about $6.5 \cdot 10^9 M_\odot$.

According to my rationale, regions of lower neutrino energy density produced Fermi seas at later times, but with larger masses. The consequences were larger supermassive black holes. 
Multiplying Eq.\,(\ref{eq:Omegat}) by the volume of a sphere with the Schwarzschild radius $r_{{\rm S},{\rm pBH}} = r_{{\rm S},\odot} M_{\rm pBH} / M_\odot $ shows that the mass of a primordial supermassive black hole is proportional to the time when it emerged, given by
\vspace{2mm}
\begin{equation}
\label{eq:t}
t_{\rm pBH} = t_{\nu {\rm d}} \cdot \frac{M_{\rm pBH}}{M_\odot} \cdot \sqrt{(\Omega_{pp} + \Omega_{\nu {\rm d}}) \frac{4\pi}{3} \frac{r_{{\rm S},\odot}^3}{M_\odot}}
\, \approx \, 2 \cdot 10^{-5}\,{\rm s}  \cdot \frac{M_{\rm pBH}}{M_\odot}\, .
\end{equation} 
\vspace{5mm}

The physics of supermassive-black-hole emergence outlined here allows for two predictions. The first prediction is that there are no supermassive black holes with masses below 40,000 solar masses. This is in line with what we know today from astronomical observations.
The second prediction is that there are no correlations between supermassive black hole masses and their DM halo masses. This prediction is in perfect agreement to the discovery that indeed such correlations do not exist \cite{Kormendy2011}. My hypothesis is in fact able to explain their discovery. 
The second prediction arises because, according to the physics outlined above, primordial black holes are \emph{not} the result of gravitational mass agglomeration. 
Supermassive black holes are simply parts of large neutrino Fermi seas. The spectrum of their sizes originated from quantum fluctuations, which are random fluctuations. Whether Fermi seas got superimposed with already formed supermassive black holes was also random.

The emergence of supermassive black holes without gravitational collapse is an interesting physical scenario. In the instant when the event horizon is physically defined, the enclosed space-time is of the same archetype as the outside. From then on, however, the space-time beyond the horizon needs to be seen as an independent universe with rather low total energy, which follows its own independent evolution.

\section{Dark matter halos and the neutrino rest mass} \label{sec:mass}

Astronomical observations suggest that every spiral galaxy contains a supermassive black hole surrounded by a DM halo. The mass of the latter is usually orders of magnitude greater than that of the black hole. If my hypothesis is correct, both supermassive BHs as well as DM halos have evolved as a result of neutrino decoupling. Supermassive black holes that initially were independent from DM distributions got either surrounded by a DM Fermi sea since the latter continued to expand, or evolved at later times, when inhomogeneities in the energy density (cosmic voids) had evolved, to supermassive\;BH/DM\;halo systems under the emission of gravitational waves \cite{GW150914}.

The formation of a supermassive\;BH/DM\;halo system can be understood similarly to the formation of a hydrogen atom from a proton and a mode of the electron field that is populated by an electron and whose extension in spacetime is large. The localization of the electron in the vicinity of the proton enforces an increase of the electrons momentum uncertainty. The increased kinetic energy balances the reduced potential of the electron. The electron remains in a pure, zero temperature state. No electro-magnetic energy is radiated away, if the process of electron localization is spherical. 
A supermassive black hole together with a dark matter halo can be seen as a giant gravitationally-bound atom. 
Different from true atoms, whose positively charged core is (partially) compensated by a trapped electron, the mass of the black hole is not compensated by a trapped neutrino field. The neutrino field rather expands the attractive $1/r$-potential to a larger and less steep one. \\
Similar to electron ground-state orbitals in hydrogen atoms, DM halo cores should have close to rotationally symmetric shapes. The DM halo of our galaxy has a triaxial shape \cite{Law2009} that is not far off a sphere. Deformations might be explainable by a significant spin of the central black hole, by the partial overlap with nearby DM halos, as well as the distribution of other masses such as smaller black holes and ordinary baryonic mass. 

Following my hypothesis, the energy density of the DM halo of the Milky Way corresponds to the product of the average galactic neutrino number density and the neutrino energy. According to \cite{Nesti2013}, the Milky Way's energy density accounts to $1.5\,{\rm GeV}/{\rm cm}^3$ in the halo core. To provide this energy density in terms of a neutrino field, the effective neutrino red shift in the Milky Way according to Eq.\,(\ref{eq:Omega}) is $z^{\nu{\rm MW}}_{\nu {\rm d}} \approx 5 \cdot 10^7$. 
If the neutrinos of the Milky Way's DM halo core indeed experienced a red shift smaller than that of photons, their number density scales up accordingly with respect to the literature value of today's neutrino density of $3.4 \cdot 10^2 / {\rm cm}^3$ \cite{Sciama1994,Pastor2011}.  
It translates to a neutrino number density in the Milky Way halo core of $(\tilde{z}^\nu_{\nu {\rm d}} / z^{\nu{\rm MW}}_{\nu {\rm d}})^3 \times 3.4 \cdot 10^2 / {\rm cm}^3 \approx 10^{10} / {\rm cm}^3$ and to an average (relativistic) neutrino energy of about $3 \cdot 2.6\,{\rm MeV} / (5 \cdot 10^7) \approx 0.15$\,eV.\\

If galactic DM halos are neutrino-pair fields that are gravitationally allocated around supermassive black holes, there is a lower bound to the radius of a DM halo. 
This lower bound corresponds to the position uncertainty of a single neutrino pair when trapped by a black hole. 
For a single neutrino pair, the gravitational potential is harmonic, and according to quantum theory the ground-state position uncertainty of the particle around the BH is given by
\begin{equation}\label{eq:dx}
\Delta \hat x = \sqrt{\frac{\hbar}{2\,m\, \omega}} \, ,
\end{equation}
where $\omega = E_{\rm p} / \hbar$ is the trapping frequency and $E_{\rm p}$  the trapping potential for the particle. 
$E_{\rm p}$ corresponds to the energy that is added to the halo if a single particle that has zero average momentum with respect to the BH at infinite distance and a quantum delocalization identical to that in the halo falls into the BH, i.e.~$E_{\rm p} =  m c^2$, yielding
\begin{equation}\label{eq:x}
\Delta \hat x = \frac{\hbar}{\sqrt{2} \,m\, c}  \, .
\end{equation}
The above equation shows that the neutrino-pair mass $m$ could be calculated if the lower bound of DM halo radii was known. 
For completeness, the momentum uncertainty is given by $\Delta \hat p = \sqrt{\hbar \omega m /2} = m c / \sqrt{2}$, fulfilling Heisenberg's uncertainty principle. 
The expectation values of the particle's position $\langle x \rangle$ and momentum $\langle p \rangle$, as defined in the BH's inertial frame and with respect to the BH's event horizon, are zero. 
Finally, the zero point energy of the (empty) DM halo is given by $E_{\rm{zpe}} = \hbar \omega/2 = \Delta^2 p /(2m) + m \omega^2 \Delta^2 x/2  = m c^2/2$,
where $\Delta^2$ denotes the variance of the observable's quantum uncertainty.

In reality, masses of galactic DM halos are several orders of magnitude larger than the masses of the central black holes and their core radii should be several orders of magnitude larger than $\Delta \hat x$ in Eq.\,(\ref{eq:x}) because an extended mass distribution allows for a larger $\Delta \hat x$ without increasing the potential energy. Heisenberg's uncertainty relation thus offers a smaller value for $\Delta \hat p$, resulting in reduced kinetic energy $\propto \Delta^2 p$ and total energy.

Not only spiral galaxies have a dark matter halo and a supermassive black hole, but also dwarf galaxies \cite{Reines2013}. An example is the Ursa Minor  spheroidal dwarf galaxy, which is a satellite galaxy of the Milky Way and which has a dark matter halo core of 0.1\,kpc radius \cite{Boyarsky2007}. 
It is a DM dominated galaxy, with a halo mass several orders of magnitude larger than the mass of its black hole. The smallness of its halo, nevertheless, provides an interesting, albeit very conservative lower bound for the lightest neutrino mass $m_3$.
Applying $\Delta \hat x_{\rm DM} = 0.1$\,kpc $\approx 3 \cdot 10^{18}$\,m to Eq.\,(\ref{eq:x}) yields
\begin{equation*}\label{eq:m}
m_3 \ggg \frac{\hbar}{2\sqrt{2} \cdot c \cdot \Delta \hat x_{\rm{DM}}}  \approx 4 \cdot 10^{-62} \,{\rm kg} \approx 2 \cdot 10^{-26} {\rm eV}\!/c^2 \, .
\end{equation*}
The actual value for $m_3$ certainly is several orders of magnitude larger than this lower bound. 
The standard model of particle physics makes the (wrong) prediction that the neutrino has \emph{zero} rest mass. Based on this, it seems not unlikely that one of the three neutrino rest masses is much lower than the other two, i.e.~$m_1 \approx 49.5$\,meV, $m_2 \approx 48.7$\,meV, and $m_3 \lll \,$1\,meV.
Modeling of the spatial distribution of dark matter in galaxies and galaxy clusters related to Eq.\,(\ref{eq:x}) but taking into account DM mass as well as other mass distributions should yield the actual value for $m_3$.

\section{Galaxy clusters and structure formation} 
Galaxy clusters contain hundreds or even thousands of galaxies. A specific feature of such clusters is that they do not expand with the general expansion of space time, similar to individual galaxies. In contrast, some `super clusters' do expand with the universe \cite{Chon2015}.
So far, it is assumed that clusters are held together by solely the gravitational force.   
Based on the nature of dark matter proposed here, I conclude that cluster formation as well as cluster internal dynamics have been strongly influenced by a quantum mechanical effect -- a binding energy between the supermassive blackholes in different galaxies due to joint DM halos. This binding energy results in a so-called covalent bond, whose fundamentals are known. 
Consider two neighboring DM-halo/BH-systems whose halos partially overlap. Such a joint DM halo represents a joint orbital excited by a joint neutrino field.
The size of the joint orbital is elongated along the symmetry axis by a factor of order two. In comparison to two DM-halo/BH-systems at large distance, the position uncertainty of the dark matter $\Delta \hat x$ is increased (without an increase in potential energy). Consequently, Heisenberg's uncertainty principle allows for the reduction of the momentum uncertainty $\Delta \hat p_x$. The neutrino field's kinetic and total energy are reduced as well, which establish the covalent bond.
The proposed covalent bond corresponds to the covalent bond in molecules, such as in hydrogen or oxygen molecules, as outlined in Refs.\,\cite{Ruedenberg1962,Baird1986,Nordholm1988}. (Unfortunately, the physical description of the covalent bond on the basis of quantum uncertainties ist still not standard in relevant textbooks.)
From this point of view, galaxy clusters could be seen as cosmological molecules. 
Different from molecules, there is no repelling force between the nuclei (the supermassive black holes), since gravitation is always attractive. The repelling mechanism, against which the bond balances, is the expansion of the universe. 
The `gravito-covalent' bonds between galaxies stabilise galaxy clusters against merging of galaxies, because merging would reduce $\Delta \hat x$ of the joint orbital, which requires energy that is not available. This quantum mechanical mechanism is identical to the one that prevents galactic dark matter halos falling into the central black hole. 
The quantum mechanical binding energy between galaxies may correspond to a significant fraction of the total mass of joint DM halos. It should then significantly contribute to the stability of galaxy clusters against both, the expansion of the universe as well as merging of supermassive BHs.

My hypothesis might shine new light on the analysis of collisions of galaxy clusters. In such collisions, three different ingredients have to be considered: stars, intra-cluster gas, and dark matter. Stars are well-spaced and their motion is only influenced by gravity. In contrast, intra-cluster gas has a large cross-section, is slowed down during collision, and shows X-ray emission. This effect is clearly observed in the bullet cluster (1E 0657-558) \cite{Clowe2006}, which is a collision of two galaxy clusters that happened 150 million years ago. If DM consisted of weakly-interacting massive particles \cite{Bertone2018}, DM should exactly accelerate as stars. If my hypothesis is correct and dark matter is a quantum field, however, its redistribution should be influenced by interference as well as the covalent binding force.

\section{Summary}
The decoupling of the neutrinos from the rest of the universe happened about half a second after the Big Bang, which is a known fact.
The prevailing description of neutrino decoupling gives a physical picture of neutrinos with trajectories, i.e.~neutrinos that move like classical particles. I argue that such a particle-like picture is incorrect, even at times before decoupling.\\
After decoupling, according to my rationale, the neutrino wave-functions expanded with almost the speed of light, overlapped, and evolved into macroscopic and massive neutrino fields. I provide an expression for the temperature of a neutrino field, according to which a field of radius $r$ has the same temperature as a Schwarzschild black hole of about the same radius. I show that the neutrino fields evolved into their quantum mechanical ground states, so-called Fermi seas.
Following the BSC theory, I argue that the neutrinos in a Fermi sea paired to bosons of zero spin, similar to Cooper pairing of electrons in superconducting metals. \\
The nonclassical feature of the Fermi seas prohibited any kind of density fluctuation, whose direct consequence was the emergence of primordial supermassive black holes without gravitational collapse. 
My hypothesis suggests a direct link between neutrino decoupling time and the mass of the smallest supermassive black holes. For the actual decoupling time of half a second after the Big Bang, my hypothesis predicts a lower bound of about $4 \cdot 10^4$ solar masses. This value is in excellent agreement with the smallest known supermassive black hole ($5 \cdot 10^4\,M_\odot$) \cite{Baldassare2015}. I show that the small exemplars evolved as early as one second after the Big Bang. Larger ones emerged in regions of smaller quantum field density, but over larger volumes and at later times. 
Subsequently, supermassive black holes and remaining dark-matter fields formed the dark component of todays galaxies and galaxy clusters.
My hypothesis predicts zero correlations between the sizes of supermassive black holes and masses of their dark matter halos, which is in agreement with recent observations \cite{Kormendy2011}. \\
I argue that dark-matter fields around supermassive black holes experienced a much lower average expansion of space-time than electromagnetic radiation, since their gravitational attraction locally resisted against the expansion. 
According to my estimation, neutrino fields can explain all cold dark matter in the universe if the red-shift they experienced after decoupling was just $z^\nu_{\nu {\rm d}} \approx 2 \cdot 10^{6}$ instead of $\tilde{z}^\nu_{\nu {\rm d}} \approx 1.5 \cdot 10^{10}$, where the latter is the red-shift of relativistic cosmic-background neutrinos if they were homogeneously distributed in space-time.\\  
My `neutrino-field hypothesis' answers the long-standing question why the centres of dark matter halos have the shapes of cores rather than `cusps' \cite{Blok2010} in the following way. Similar to the electron probability distribution in a hydrogen atom in ground state, quantum position uncertainties generally have a rather flat shape at its centre. 
My hypothesis also answers the question why DM halos are stable since their formation. Again it is useful to compare a DM halo around a supermassive black hole with the electron ground-state orbital around a proton. The electron does not fall into the proton because a more precisely determined position uncertainty would request a larger momentum uncertainty and thus a larger kinetic energy.  
My hypothesis also provides a solution for the so far unknown mechanism of primordial formation of super-massive black holes.
And finally it suggests a gravito-covalent bond due to joint DM orbitals as a significant contribution to galaxy-cluster formation and stability.\\
My results and predictions as well as my entire hypothesis are based on the fact that a non-interacting particle has a wave function whose position uncertainty continuously increases. Many of such particles from the same species therefor must evolve into indistinguishable (`degenerate') particles (where the term `particle' is in fact rather inappropriate). In quantum field physics, `indistinguishable particles' are indistinguishable with respect to all their degrees of freedom, i.e.~they all join the same position. Their analysis requires statistics that are very different from the statistics of particles with distinguishable trajectories. Of course, other decoupled particles must also be analyzed using the statistics of the indistinguishable particles. This includes proposed as yet undiscovered particles. Examples are the hypothetical \emph{weakly-interacting massive particles} (WIMPs) and \emph{weakly-interacting slim particles} (WISPs) \cite{Goodman1985,Jungman1996,Ringwald2012,Klasen2015,Bertone2018}. 
The prevailing semiclassical, particle-like statistical view of WIMPS and WISPS does not correspond to the fact of quantum field physics that, without interactions, positional uncertainties become large and thus interference becomes relevant. \\

\subsection*{Acknowledgements}
This work was funded by the European Research Council (ERC) project `MassQ' (Grant No.~339897).
The author acknowledges Wilfried Buchm\"uller, Ludwig Mathey, Henning Moritz for helpful discussions, and Mikhail Korobko and Christian Rembe for valuable comments on the manuscript.
The author further acknowledges useful discussion within `Quantum Universe' (Grant No.~390833306), which is financed by the Deutsche Forschungsgemeinschaft (DFG, German Research Foundation) under Germany's Excellence Strategy - EXC 2121.


\section{References}

\begin{thebibliography}{10}

\bibitem{Massey2010}
R.~Massey, T.~Kitching, and J.~Richard, ``{The dark matter of gravitational
  lensing},'' {\em Reports on Progress in Physics}, vol.~73, p.~086901, aug
  2010.

\bibitem{DeSwart2017}
J.~G. de~Swart, G.~Bertone, and J.~van Dongen, ``{How dark matter came to
  matter},'' {\em Nature Astronomy}, vol.~1, p.~0059, mar 2017.

\bibitem{Rubin1970}
V.~C. Rubin and W.~K. {Ford Jr.}, ``{Rotation of the Andromeda Nebula from a
  Spectroscopic Survey of Emission Regions},'' {\em The Astrophysical Journal},
  vol.~159, p.~379, feb 1970.

\bibitem{Ostriker1973}
J.~P. Ostriker and P.~J.~E. Peebles, ``{A Numerical Study of the Stability of
  Flattened Galaxies: or, can Cold Galaxies Survive?},'' {\em The Astrophysical
  Journal}, vol.~186, p.~467, dec 1973.

\bibitem{Einasto1974}
J.~Einasto, A.~Kaasik, and E.~Saar, ``{Dynamic evidence on massive coronas of
  galaxies},'' {\em Nature}, vol.~250, pp.~309--310, jul 1974.

\bibitem{Rubin1978}
V.~C. Rubin, N.~Thonnard, and J.~{Ford, W. K.}, ``{Extended rotation curves of
  high-luminosity spiral galaxies. IV - Systematic dynamical properties, SA
  through SC},'' {\em The Astrophysical Journal}, vol.~225, p.~L107, nov 1978.

\bibitem{Rubin1980}
V.~C. Rubin, N.~Thonnard, and J.~{Ford, W. K.}, ``{Rotational properties of 21
  SC galaxies with a large range of luminosities and radii, from NGC 4605 (R =
  4kpc) to UGC 2885 (R = 122 kpc)},'' {\em The Astrophysical Journal},
  vol.~238, p.~471, jun 1980.

\bibitem{Kafle2014}
P.~R. Kafle, S.~Sharma, G.~F. Lewis, and J.~Bland-Hawthorn, ``{On the shoulders
  of giants: Properties of the stellar halo and the milky way mass
  distribution},'' {\em Astrophysical Journal}, vol.~794, no.~1, 2014.

\bibitem{Mellier1999}
Y.~Mellier, ``{Probing the Universe with Weak Lensing},'' {\em Annual Review of
  Astronomy and Astrophysics}, vol.~37, pp.~127--189, sep 1999.

\bibitem{Clowe2006}
D.~Clowe, M.~Brada{\v{c}}, A.~H. Gonzalez, M.~Markevitch, S.~W. Randall,
  C.~Jones, and D.~Zaritsky, ``{A Direct Empirical Proof of the Existence of
  Dark Matter},'' {\em The Astrophysical Journal}, vol.~648, pp.~L109--L113,
  sep 2006.

\bibitem{Smoot1992}
G.~F. Smoot, C.~L. Bennett, A.~Kogut, E.~L. Wright, J.~Aymon, N.~W. Boggess,
  E.~S. Cheng, G.~de~Amici, S.~Gulkis, M.~G. Hauser, G.~Hinshaw, P.~D. Jackson,
  M.~Janssen, E.~Kaita, T.~Kelsall, P.~Keegstra, C.~Lineweaver, K.~Loewenstein,
  P.~Lubin, J.~Mather, S.~S. Meyer, S.~H. Moseley, T.~Murdock, L.~Rokke, R.~F.
  Silverberg, L.~Tenorio, R.~Weiss, and D.~T. Wilkinson, ``{Structure in the
  COBE differential microwave radiometer first-year maps},'' {\em The
  Astrophysical Journal}, vol.~396, p.~L1, sep 1992.

\bibitem{Hinshaw2009}
G.~Hinshaw, J.~L. Weiland, R.~S. Hill, N.~Odegard, D.~Larson, C.~L. Bennett,
  J.~Dunkley, B.~Gold, M.~R. Greason, N.~Jarosik, E.~Komatsu, M.~R. Nolta,
  L.~Page, D.~N. Spergel, E.~Wollack, M.~Halpern, A.~Kogut, M.~Limon, S.~S.
  Meyer, G.~S. Tucker, and E.~L. Wright, ``{Five-year Wilkinson microwave
  anisotropy probe observations: data processing, sky maps, and basic
  results},'' {\em The Astrophysical Journal Supplement Series}, vol.~180,
  pp.~225--245, feb 2009.

\bibitem{Ade2016}
P.~A.~R. Ade~\emph{et al.}, ``{Planck 2015 results},'' {\em Astronomy {\&}
  Astrophysics}, vol.~594, p.~A13, oct 2016.

\bibitem{Spergel2015}
D.~N. Spergel, ``{The dark side of cosmology: Dark matter and dark energy},''
  {\em Science}, vol.~347, pp.~1100--1102, mar 2015.

\bibitem{Pauli1925}
W.~Pauli, ``{{\"{U}}ber den Zusammenhang des Abschlusses der Elektronengruppen
  im Atom mit der Komplexstruktur der Spektren},'' {\em Zeitschrift f{\"{u}}r
  Physik}, vol.~31, pp.~765--783, feb 1925.

\bibitem{Tremaine1979}
S.~Tremaine and J.~E. Gunn, ``{Dynamical Role of Light Neutral Leptons in
  Cosmology},'' {\em Physical Review Letters}, vol.~42, pp.~407--410, feb 1979.

\bibitem{White1983}
S.~D.~M. White, C.~S. Frenk, and M.~Davis, ``{Clustering in a
  neutrino-dominated universe},'' {\em The Astrophysical Journal}, vol.~274,
  p.~L1, nov 1983.

\bibitem{Goodman1985}
M.~W. Goodman and E.~Witten, ``{Detectability of certain dark-matter
  candidates},'' {\em Physical Review D}, vol.~31, pp.~3059--3063, jun 1985.

\bibitem{Jungman1996}
G.~Jungman, M.~Kamionkowski, and K.~Griest, ``{Supersymmetric dark matter},''
  {\em Physics Reports}, vol.~267, pp.~195--373, mar 1996.

\bibitem{Ringwald2012}
A.~Ringwald, ``{Exploring the role of axions and other WISPs in the dark
  universe},'' {\em Physics of the Dark Universe}, vol.~1, pp.~116--135, nov
  2012.

\bibitem{Klasen2015}
M.~Klasen, M.~Pohl, and G.~Sigl, ``{Indirect and direct search for dark
  matter},'' {\em Progress in Particle and Nuclear Physics}, vol.~85,
  pp.~1--32, nov 2015.

\bibitem{Bertone2018}
G.~Bertone and T.~M.~P. Tait, ``{A new era in the search for dark matter},''
  {\em Nature}, vol.~562, pp.~51--56, oct 2018.

\bibitem{deSalas2016}
P.~F. de~Salas and S.~Pastor, ``{Relic neutrino decoupling with flavour
  oscillations revisited},'' {\em Journal of Cosmology and Astroparticle
  Physics}, vol.~2016, pp.~051--051, jul 2016.

\bibitem{Sciama1994}
{D. W. Sciama}, {\em {Modern Cosmology and the Dark Matter Problem}}.
\newblock Cambridge University Press, 1994.

\bibitem{EXO-200-2014}
{The EXO-200 collaboration}, ``{Search for Majorana neutrinos with the first
  two years of EXO-200 data},'' {\em Nature}, vol.~510, pp.~229--234, jun 2014.

\bibitem{Liu2012}
F.-H. Liu, C.-X. Tian, M.-Y. Duan, and B.-C. Li, ``{Relativistic and Quantum
  Revisions of the Multisource Thermal Model in High-Energy Collisions},'' {\em
  Advances in High Energy Physics}, vol.~2012, pp.~1--9, 2012.
  
\bibitem{Heisenberg1927}
W.~Heisenberg, {\em {\"{U}}ber den anschaulichen Inhalt der quantentheoretischen Kinematik und Mechanik}, 
{Z. Phys} 43, 172--198 (1927).

\bibitem{Kennard1927}
E.~H. Kennard, {\em Zur Quantenmechanik einfacher Bewegungstypen}, 
{Z. Phys} 44, 326--352 (1927).

\bibitem{Weyl1927}
H.~Weyl, {\em Quantenmechanik und Gruppentheorie}, 
{Z. Phys} 46, 1--46 (1927).

\bibitem{Robertson1929}
H.~P. Robertson, {\em The Uncertainty Principle}, 
{Phys. Rev.} 34, 163--164 (1929).

\bibitem{Pastor2011}
S.~Pastor, ``{Light neutrinos in cosmology},'' {\em Physics of Particles and
  Nuclei}, vol.~42, no.~4, pp.~628--640, 2011.

\bibitem{Aspect1981}
A.~Aspect, P.~Grangier, and G.~Roger, ``{Experimental Tests of Realistic Local
  Theories via Bell`s Theorem},'' {\em Physical Review Letters}, vol.~47,
  p.~460, 1981.

\bibitem{Schnabel2020}
R.~Schnabel, ``{\;`Quantum weirdness' in exploitation by the international
  gravitational-wave observatory network},'' {\em Annalen der Physik},
  vol.~532, p.~1900508, mar 2020.

\bibitem{Planck1900}
M.~Planck, ``{{\"{U}}ber das Gesetz der Energieverteilung im Normalspektrum},''
  {\em Annalen der Physik}, vol.~4, no.~4, pp.~553--563, 1900.

\bibitem{Hawking1974}
S.~W. Hawking, ``{Black hole explosions?},'' {\em Nature}, vol.~248,
  pp.~30--31, mar 1974.
  
\bibitem{Schwarzschild1916b}
K. Hawking, ``{{\"{U}}ber das Gravitationsfeld einer Kugel aus inkompressibler Fl{\"{u}}ssigkeit nach der Einsteinschen Theorie},'' {\em Sitzungsberichte der K{\"{o}}niglich Preu{\ss}ischen Akademie der Wissenschaften}, 424-434, mar 1916. 

\bibitem{Cuesta2016}
A.~J. Cuesta, V.~Niro, and L.~Verde, ``{Neutrino mass limits: Robust
  information from the power spectrum of galaxy surveys},'' {\em Physics of the
  Dark Universe}, vol.~13, pp.~77--86, sep 2016.

\bibitem{Capozzi2016}
F.~Capozzi, E.~Lisi, A.~Marrone, D.~Montanino, and A.~Palazzo, ``{Neutrino
  masses and mixings: Status of known and unknown 3 $\nu$ parameters},'' {\em
  Nuclear Physics B}, vol.~908, pp.~218--234, jul 2016.

\bibitem{Bardeen1957}
J.~Bardeen, L.~N. Cooper, and J.~R. Schrieffer, ``{Theory of
  Superconductivity},'' {\em Physical Review}, vol.~108, pp.~1175--1204, dec
  1957.

\bibitem{Cooper1956}
L.~N. Cooper, ``{Bound Electron Pairs in a Degenerate Fermi Gas},'' {\em
  Physical Review}, vol.~104, pp.~1189--1190, nov 1956.

\bibitem{Einstein1935}
A.~Einstein, B.~Podolsky, and N.~Rosen, ``{Can Quantum-Mechanical Description
  of Physical Reality Be Considered Complete?},'' {\em Physical Review},
  vol.~47, pp.~777--780, may 1935.

\bibitem{Schnabel2015}
R.~Schnabel, ``{Einstein-Podolsky-Rosen--entangled motion of two massive
  objects},'' {\em Physical Review A}, vol.~92, p.~012126, jul 2015.

\bibitem{Geller1989}
M.~J. Geller and J.~P. Huchra, ``{Mapping the Universe},'' {\em Science},
  vol.~246, pp.~897--903, nov 1989.

\bibitem{Palanque-Delabrouille2015}
N.~Palanque-Delabrouille, C.~Y{\`{e}}che, J.~Baur, C.~Magneville, G.~Rossi,
  J.~Lesgourgues, A.~Borde, E.~Burtin, J.-M. LeGoff, J.~Rich, M.~Viel, and
  D.~Weinberg, ``{Neutrino masses and cosmology with Lyman-alpha forest power
  spectrum},'' {\em Journal of Cosmology and Astroparticle Physics}, vol.~2015,
  pp.~011--011, nov 2015.

\bibitem{Komatsu2011}
E.~Komatsu, K.~M. Smith, J.~Dunkley, C.~L. Bennett, B.~Gold, G.~Hinshaw,
  N.~Jarosik, D.~Larson, M.~R. Nolta, L.~Page, D.~N. Spergel, M.~Halpern, R.~S.
  Hill, A.~Kogut, M.~Limon, S.~S. Meyer, N.~Odegard, G.~S. Tucker, J.~L.
  Weiland, E.~Wollack, and E.~L. Wright, ``{SEVEN-YEAR WILKINSON MICROWAVE
  ANISOTROPY PROBE ( WMAP ) OBSERVATIONS: COSMOLOGICAL INTERPRETATION},'' {\em
  The Astrophysical Journal Supplement Series}, vol.~192, p.~18, feb 2011.

\bibitem{Einstein1905srt}
A.~Einstein, ``{Zur Elektrodynamik bewegter K{\"{o}}rper},'' {\em Annalen der
  Physik}, vol.~322, no.~10, pp.~891--921, 1905.

\bibitem{Baldassare2015}
V.~F. Baldassare, A.~E. Reines, E.~Gallo, and J.~E. Greene, ``{A $\sim
  \!50,000$ M$_\odot$ solar mass black hole in the nucleus of RGG 118},'' {\em
  The Astrophysical Journal}, vol.~809, p.~L14, aug 2015.

\bibitem{Akiyama2019}
{The Event Horizon Telescope Collaboration}, ``{First M87 Event Horizon
  Telescope Results. VI. The Shadow and Mass of the Central Black Hole},'' {\em
  The Astrophysical Journal}, vol.~875, p.~L6, apr 2019.

\bibitem{Kormendy2011}
J.~Kormendy and R.~Bender, ``{Supermassive black holes do not correlate with
  dark matter haloes of galaxies},'' {\em Nature}, vol.~469, pp.~377--380, jan
  2011.

\bibitem{GW150914}
B.~P. Abbott~\emph{et al.}, ``{Observation of Gravitational Waves from a Binary
  Black Hole Merger},'' {\em Physical Review Letters}, vol.~116, p.~061102, feb
  2016.

\bibitem{Law2009}
D.~R. Law, S.~R. Majewski, and K.~V. Johnston, ``{Evidence for a triaxial Milky
  Way dark matter halo from the Sagittarius stellar tidal stream},'' {\em The
  Astrophysical Journal}, vol.~703, pp.~L67--L71, sep 2009.

\bibitem{Nesti2013}
F.~Nesti and P.~Salucci, ``{The Dark Matter halo of the Milky Way, AD 2013},''
  {\em Journal of Cosmology and Astroparticle Physics}, vol.~2013,
  pp.~016--016, jul 2013.

\bibitem{Reines2013}
A.~E. Reines, J.~E. Greene, and M.~Geha, ``{Dwarf galaxies with optical
  signatures of active massive black holes},'' {\em The Astrophysical Journal},
  vol.~775, p.~116, sep 2013.

\bibitem{Boyarsky2007}
A.~Boyarsky, J.~Nevalainen, and O.~Ruchayskiy, ``{Constraints on the parameters
  of radiatively decaying dark matter from the dark matter halos of the Milky
  Way and Ursa Minor},'' {\em Astronomy {\&} Astrophysics}, vol.~471,
  pp.~51--57, aug 2007.

\bibitem{Chon2015}
G.~Chon, H.~B{\"{o}}hringer, and S.~Zaroubi, ``{On the definition of
  superclusters},'' {\em Astronomy {\&} Astrophysics}, vol.~575, p.~L14, mar
  2015.

\bibitem{Ruedenberg1962}
K.~Ruedenberg, ``{The Physical Nature of the Chemical Bond},'' {\em Reviews of
  Modern Physics}, vol.~34, pp.~326--376, apr 1962.

\bibitem{Baird1986}
N.~C. Baird, ``{The chemical bond revisited},'' {\em Journal of Chemical
  Education}, vol.~63, p.~660, aug 1986.

\bibitem{Nordholm1988}
S.~Nordholm, ``{Delocalization -- the key concept of covalent bonding},'' {\em
  Journal of Chemical Education}, vol.~65, p.~581, jul 1988.

\bibitem{Blok2010}
W.~J.~G. de~Blok, ``{The Core-Cusp Problem},'' {\em Advances in Astronomy},
  vol.~2010, pp.~1--14, 2010.

\end{thebibliography}
\end{document}